\documentstyle[prd,aps,floats,tighten,epsfig]{revtex}

\newcommand{\hepph}[1]{{\tt hep-ph/#1}}
\newcommand{\astroph}[1]{{\tt astro-ph/#1}}
\newcommand{\prep}[3]{Phys.\ Rep.\ {\bf #1} (#2) #3}
\newcommand{\plb}[3]{Phys.\ Lett.\ {\bf B#1} (#2) #3}
\newcommand{\npb}[3]{Nucl.\ Phys.\ {\bf B#1} (#2) #3}
\newcommand{\cpc}[3]{Comm.\ Phys.\ Comm.\ {\bf #1} (#2) #3}
\renewcommand{\apj}[3]{Astrophys.\ J.\ {\bf #1} (#2) #3}
\renewcommand{\prl}[3]{Phys.\ Rev.\ Lett. {\bf #1} (#2) #3}
\renewcommand{\prd}[3]{Phys.\ Rev.\ {\bf D#1} (#2) #3}
\renewcommand{\rmp}[3]{Rev.\ Mod.\ Phys.\ {\bf #1} (#2) #3}
\newcommand{\href}[2]{#1}
\newcommand{\email}[1]{\tt #1}
\newcommand{\sigv}{\langle\sigma v\rangle_{\rm tot}}
\newcommand{\taue}{\tau_E}
\newcommand{\taud}{\tau_D}
\newcommand{\deltav}{\Delta v}
\newcommand{\mct}{{\tilde{m}_\chi}}
\def\erf{\mathop{\rm erf}}

\begin{document}
	
\draft




\title{Positron Propagation and Fluxes from Neutralino Annihilation
in the Halo}

\author{Edward~ A.~ Baltz}
\address{Department of Physics, University of California, Berkeley, CA
94720, USA; \\
E-mail: \email{eabaltz@astron.berkeley.edu}}

\author{Joakim Edsj\"o}
\address{Center for Particle Astrophysics, University of California,
301 Le Conte Hall, Berkeley, \\
CA 94720-7304, USA; 
E-mail: \email{edsjo@cfpa.berkeley.edu}}

\maketitle


\begin{abstract}
Supersymmetric neutralinos are one of the most promising candidates 
for the dark matter in the Universe.  If they exist, they should make 
up some fraction of the Milky Way halo.  We investigate the fluxes of 
positrons expected at the Earth from neutralino annihilation in the 
halo.  Positron propagation is treated in a diffusion model including 
energy loss.  The positron source function includes contributions from 
both continuum and monochromatic positrons.  We find that, for a 
``canonical'' halo model and propagation parameters, the fluxes are 
generally too low to be visible.  Given the large uncertainties in 
both propagation and halo structure, it is however possible to obtain 
observable fluxes.  We also investigate the shapes of the positron 
spectra, including fits to a feature indicated by the results of the 
{\sc Heat} experiment.
\end{abstract}

\pacs{95.35.+d, 14.80.Ly, 98.70.Sa}






\section{Introduction} \label{sec:intro}

It is well known that a large fraction of the mass of the Universe has
only been observed by its gravitational effects.  The nature of this
dark matter is unknown.  Standard Big Bang Nucleosynthesis bounds the
baryon density of the Universe, but these results fall far short of
the known amount of dark matter.  Thus more exotic forms of matter are
sought, and weakly interacting massive particles (WIMPs) are one of
the most promising cold dark matter candidates.

Perhaps the best WIMP candidate is the neutralino (for a review, see
Ref.\ \cite{jkg}), which arises in supersymmetric models as a linear
combination of the superpartners of the neutral gauge and Higgs
bosons.  We choose the neutralino as our WIMP candidate, though we
allow its properties to vary over a generous sample of supersymmetric
models.

It may be possible to detect neutralinos (or WIMPs) in the galactic
halo by the products of their mutual annihilations, e.g.\ by searching
for gamma rays \cite{gammahalo}, antiprotons \cite{pbarhalo} and
positrons \cite{epcont,epline,kamturner} coming from the Milky Way
halo.  In this paper we consider neutralino annihilation in the halo
into positrons (both continuum and monochromatic).  Compared to
earlier studies \cite{epcont,epline,kamturner}, we will use a true
diffusion model instead of a leaky-box model.  This has the advantage
of allowing us to use more realistic models for the diffusion constant
while still permitting an analytic solution.  The numerical
implementation is very fast, making detailed scans over the huge MSSM
parameter space feasible.  We also include all two-body final states
of neutralino annihilation (at tree level) using detailed Monte Carlo
simulations for the decay and/or hadronization of the annihilation
products.  We will compare our results mainly with those of
Kamionkowski and Turner \cite{kamturner} (hereafter KT).

We will also compare our predictions of the positron fluxes to
observations. The fluxes are typically small compared to the expected
background, but given the large uncertainties involved, we identify
models that can have a significant effect on the observed positron
spectrum.  We will show that a large class of models improves the fit
of observations to the predicted background.  Finally, we investigate
the positron spectra at higher energies than have been observed, and
identify features that may be detectable by future experiments such
as {\sc Ams} \cite{ams}.


\section{Set of supersymmetric models}

We work in the Minimal Supersymmetric Standard Model (MSSM)\@.  In
general, the MSSM has many free parameters, but with some reasonable
assumptions we can reduce the number of parameters to the Higgsino
mass parameter $\mu$, the gaugino mass parameter $M_{2}$, the ratio of
the Higgs vacuum expectation values $\tan \beta$, the mass of the
$CP$-odd Higgs boson $m_{A}$, the scalar mass parameter $m_{0}$ and
the trilinear soft SUSY-breaking parameters $A_{b}$ and $A_{t}$ for
the third generation.  In particular, we don't impose any restrictions
from supergravity other than gaugino mass unification.  We have made
some scans without the GUT relation for the gaugino mass parameters
$M_1$ and $M_2$\@.  This mainly has the effect of allowing lower
neutralino masses to escape the LEP bounds and is not very important
for this study.  Hence, the GUT relation for $M_1$ and $M_2$ is kept
throughout this paper.  For a more detailed definition of the
parameters and a full set of Feynman rules, see Refs.\
\cite{coann,jephd}.

The lightest stable supersymmetric particle is in most models the
lightest neutralino, which is a superposition of the superpartners of
the gauge and Higgs fields,
\begin{equation}
  \tilde{\chi}^0_1 = 
  N_{11} \tilde{B} + N_{12} \tilde{W}^3 + 
  N_{13} \tilde{H}^0_1 + N_{14} \tilde{H}^0_2.
\end{equation}
It is convenient to define the gaugino fraction of the lightest neutralino,
\begin{equation}
  Z_g = |N_{11}|^2 + |N_{12}|^2.
\end{equation}
For the masses of the neutralinos and charginos we use the one-loop
corrections as given in Ref.\ \cite{neuloop} and for the Higgs boson
masses we use the leading log two-loop radiative corrections,
calculated within the effective potential approach given in Ref.\
\cite{carena}.

We make extensive scans of the model parameter space, some general and
some specialized to interesting regions.  In total we make 18
different scans of the parameter space.  The scans are done randomly
and are mostly distributed logarithmically in the mass parameters and
in $\tan \beta$\@.  For some scans the logarithmic scan in $\mu$ and
$M_{2}$ has been replaced by a logarithmic scan in the more physical
parameters $m_{\chi}$ and $Z_{g}/(1-Z_{g})$ where $m_{\chi}$ is the
neutralino mass. Combining all the scans, the overall limits of the
seven MSSM parameters we use are given in Table~\ref{tab:scans}.

We check each model to see if it is excluded by the most recent
accelerator constraints, of which the most important ones are the LEP
bounds \cite{lepbounds} on the lightest chargino mass,
\begin{equation}
  m_{\chi_{1}^{+}} > \left\{ \begin{array}{lcl}
  91 {\rm ~GeV} & \quad , \quad & | m_{\chi_{1}^{+}} -
  m_{\chi^{0}_{1}} | 
  > 4 {\rm ~GeV} \\
  85 {\rm ~GeV} & \quad , \quad & {\rm otherwise}
  \end{array} \right.
\end{equation}
and on the lightest Higgs boson mass $m_{H_{2}^{0}}$ (which range from
72.2--88.0 GeV depending on $\sin (\beta-\alpha)$ with $\alpha$ being
the Higgs mixing angle) and the constrains from $b \to s \gamma$
\cite{cleo}.  For each allowed model we calculate the relic density of
neutralinos $\Omega_\chi h^2$, where $\Omega_\chi$ is the density in
units of the critical density and $h$ is the present Hubble constant
in units of $100$ km s$^{-1}$ Mpc$^{-1}$\@.  We use the formalism of
Ref.~\cite{GondoloGelmini} for resonant annihilations, threshold
effects, and finite widths of unstable particles and we include all
two-body tree-level annihilation channels of neutralinos.  We also
include the so-called coannihilation processes in the relic density
calculation according to the analysis of Edsj\"o and Gondolo
\cite{coann}.

Present observations favor $h=0.6\pm 0.1$, and a total matter density
$\Omega_{M}=0.3\pm 0.1$, of which baryons may contribute 0.02 to 0.08
\cite{cosmparams}.  Not to be overly restrictive, we accept
$\Omega_\chi h^2$ in the range from $0.025$ to $1$ as cosmologically
interesting.  The lower bound is somewhat arbitrary as there may be
several different components of non-baryonic dark matter, but we
demand that neutralinos are at least as abundant as required to make
up the dark halos of galaxies.  In principle, neutralinos with
$\Omega_\chi h^2<0.025$ would still be relic particles, but only
making up a small fraction of the dark matter of the Universe.  We
will only consider models with $\Omega_\chi h^2<0.025$ when discussing
the dependence of the signal on $\Omega_\chi h^2$\@.

\begin{table}
  \begin{tabular}{rrrrrrrr} 
  Parameter & $\mu$ & $M_{2}$ &
  $\tan \beta$ & $m_{A}$ & $m_{0}$ & $A_{b}/m_{0}$ & $A_{t}/m_{0}$ \\
  Unit & GeV & GeV & 1 & GeV & GeV & 1 & 1 \\ \hline 
  Min & -50000 & -50000 & 1.0  & 0     & 100   & -3 & -3 \\
  Max & 50000  & 50000  & 60.0 & 10000 & 30000 &  3 &  3 \\  
  \end{tabular} 
\caption{The ranges of parameter values used in our scans of the MSSM parameter
  space. Note that several special scans aimed at interesting regions of the
  parameter space has been performed.  In total we have generated approximately
  103000 models that are not excluded by accelerator searches.}
  \label{tab:scans}
\end{table}

\section{Positron spectra from neutralino annihilation}

When neutralinos annihilate in the galactic halo they produce quarks,
leptons, gauge bosons, Higgs bosons and gluons.  When these particles
decay and/or hadronize, they will give rise to positrons either
directly or from decaying mesons in hadron jets. We thus expect to get
both monochromatic positrons (at an energy of $m_{\chi}$) from direct
annihilation into $e^{+}e^{-}$ and continuum positrons from the other
annihilation channels. In general the branching ratio for annihilation
directly into $e^+ e^-$ is rather small, but for some classes of
models one can obtain a large enough branching ratio for the line to
be observable.

For continuum positrons, we simulate the decay and/or hadronization 
with the Lund Monte Carlo {\sc Pythia} 6.115 \cite{pythia}.  We 
perform the simulation for a set of 18 different neutralino masses 
from 10 to 5000 GeV and for the ``basic'' annihilation channels 
consisting of the heavy quarks, leptons and gauge bosons.  The 
annihilation channels containing Higgs bosons are then easily taken 
into account by allowing the Higgs bosons to decay in flight and 
averaging the produced positron flux over the decay angles.  For any 
given MSSM model, the positron spectrum is then given by
\begin{equation}
  \frac{d\phi}{d\varepsilon} = 
  \left. \frac{d\phi}{d\varepsilon} \right|_{\rm cont.} + 
  \left. \frac{d\phi}{d\varepsilon} \right|_{\rm line}
  = \sum_{F \neq e^{+}e^{-}} B_{F} 
  \left. \frac{d\phi}{d\varepsilon} \right|_{F} + B_{e^{+}e^{-}}\, 
  \delta(\varepsilon-\tilde{m}_{\chi}), 
\end{equation}
in units of $e^{+}/{\rm annihilation}$ where $\varepsilon = 
E_{e^{+}}/({\rm 1~GeV})$, $\tilde{m}_\chi = m_\chi/({\rm 1~GeV})$, 
$B_{F}$ is the branching ratio into a given final state $F$ and 
$\left.d\phi/d\varepsilon\right|_{F}$ is the spectrum of positrons 
from annihilation channel $F$\@.  We include all two-body final states 
(except the three lightest quarks which are completely negligible) at 
tree level and the $Z\gamma$ \cite{bub} and $gg$ \cite{bu} final 
states which arise at one-loop level.

\section{Propagation model}

\subsection{Propagation and the interstellar flux}

We consider a standard diffusion model for the propagation of
positrons in the galaxy.  Charged particles move under the influence
of the galactic magnetic field.  With the energies we are concerned
with, the magnetic gyroradii of the particles are quite small.
However, the magnetic field is tangled, and even with small gyroradii,
particles can jump to nearby field lines which will drastically alter
their courses.  This entire process can be modeled as a random walk,
which can be described by a diffusion equation.

Positron propagation is complicated by the fact that light particles
lose energy quickly due to inverse Compton and synchrotron processes.
Diffuse starlight and the Cosmic Microwave Background (CMB) both
contribute appreciably to the energy loss rate of high energy
electrons and positrons via inverse Compton scattering.  Electrons and
positrons also lose energy by synchrotron radiation as they spiral
around the galactic magnetic field lines.

Our detailed treatment of positron diffusion follows.  First we define
a dimensionless energy variable $\varepsilon=E/(1\;{\rm GeV})$\@.  The
standard diffusion-loss equation for the space density of cosmic rays
per unit energy, $dn/d\varepsilon$, is given by
\begin{equation}
  \frac{\partial}{\partial t}\frac{dn}{d\varepsilon}=\vec{\nabla}\cdot
  \left[K(\varepsilon,\vec{x})\vec{\nabla}\frac{dn}{d\varepsilon}\right]+
  \frac{\partial}{\partial \varepsilon}\left[b(\varepsilon,\vec{x})
  \frac{dn}{d\varepsilon}\right]+Q(\varepsilon,\vec{x}),
  \label{eq:diffloss}
\end{equation}
where $K$ is the diffusion constant, $b$ is the energy loss rate and
$Q$ is the source term.  We will consider only steady state solutions,
setting the left hand side of Eq.\ (\ref{eq:diffloss}) to zero.

We assume that the diffusion constant $K$ is constant in space
throughout a ``diffusion zone'', but it may vary with energy.  At
energies above a few GeV, we can represent the diffusion constant as a
power law in energy \cite{wlg},
\begin{equation}
  K(\varepsilon)=K_0\varepsilon^\alpha\approx
  3\times 10^{27}\varepsilon^{0.6}{\rm cm}^2\;{\rm s}^{-1}.
  \label{eq:KA}
\end{equation}
However, at energies below about 3 GeV, there is a cutoff in the
diffusion constant that can be modeled as
\begin{equation}
  K(\varepsilon)=K_0\left[C+\varepsilon^\alpha\right]\approx
  3\times 10^{27}\left[3^{0.6}+\varepsilon^{0.6}\right]
  {\rm cm}^2\;{\rm s}^{-1}.
  \label{eq:KB}
\end{equation}
We will in the following consider both of these models for the
diffusion constant, but we will focus on the second expression.  We
will refer to Eq.~(\ref{eq:KA}) as model A and Eq.~(\ref{eq:KB}) as
model B\@. The function $b(\varepsilon)$ represents the (time) rate of
energy loss.  We allow energy loss via synchrotron emission and
inverse Compton scattering.  The rms magnetic field in the diffusion
zone is about $3~\mu$G, an energy density of about 0.2 eV
cm$^{-3}$\@. We allow inverse Compton scattering on both the cosmic
microwave background and diffuse starlight, which have energy
densities of 0.3 and 0.6 eV cm$^{-3}$ respectively.  These two
processes combined give an energy loss rate \cite{energy-loss}
\begin{equation}
  b(\varepsilon)_{e^\pm}=\frac{1}{\taue}\varepsilon^2
  \approx 10^{-16}\varepsilon^2\;{\rm s}^{-1},
  \label{eq:bofe}
\end{equation}
where we have neglected the space dependence of the energy loss rate.
The function $Q$ is the source of positrons in units of cm$^{-3}$
s$^{-1}$\@.

The diffusion zone is a slab of thickness $2L$\@.  We will fix $L$ to
be 3 kpc, which fits observations of the cosmic ray flux \cite{wlg}.
We impose free escape boundary conditions, namely that the cosmic ray
density drops to zero on the surfaces of the slab, which we let be the
planes $z=\pm L$\@.  We neglect the radial boundary usually considered
in diffusion models.  This is justified when the sources of cosmic
rays are nearer than the boundary, as is usually the case with
galactic sources.  We will see that the positron flux at Earth,
especially at higher energies, mostly originates within a few kpc and
hence this approximation is well justified in our case.

We will now consider the diffusion-loss equation, 
Eq.~(\ref{eq:diffloss}), in the steady-state.  We first make a change 
of variables, $u=1/\varepsilon$, and we write the diffusion constant 
as a function of $u$, with $K(E)=K_0h(u)$, yielding
\begin{equation}
  \frac{1}{h(u)}\frac{\partial}{\partial u}\frac{dn}{du}=
  K_0\taue\nabla^2\frac{dn}{du}-\taue\left[u^2h(u)\right]^{-1}
  Q(\varepsilon(u),\vec{x}),
\end{equation}
where $\taue=10^{16}$ s is the energy-loss time introduced in
Eq.~(\ref{eq:bofe}).  We now suppose that $\alpha<1$, and we make a
further change of variables $v=\int h(u)du$\@.  We find the following
transformations for models A and B,
\begin{equation}
  v_A=\frac{1}{1-\alpha}u^{1-\alpha}\hspace{0.1in}{\rm and}\hspace{0.1in}
  v_B=Cu+\frac{1}{1-\alpha}u^{1-\alpha}.
\end{equation}
These transformations need to be inverted.  The first one can be done
analytically,
\begin{equation}
  u=\left[(1-\alpha)v_A\right]^{\frac{1}{1-\alpha}},
\end{equation}
though the second one requires a numerical inversion.  We notice that the
function of $u$ in the source term is just the Jacobian of the full
transformation,
\begin{equation}
  \frac{1}{u^2h(u)}=-\frac{d\varepsilon}{dv}=w(v).
\end{equation}
Again, in the first case this can be evaluated,
\begin{equation}
  w(v_A)=\left[(1-\alpha)v_A\right]^{-\frac{2-\alpha}{1-\alpha}},
\end{equation}
and the second requires the implicit solution of $u(v)$\@.  We put
$F(v)=dn/du(v)$ and arrive at
\begin{equation}
  \frac{\partial}{\partial v}F(v,\vec{x})=K_0\taue\nabla^2
  F(v,\vec{x})-\taue w(v)Q(\varepsilon(v),\vec{x}).
  \label{eq:diff-v}
\end{equation}

Eq.~(\ref{eq:diff-v}) is equivalent to the inhomogeneous heat equation.  The
space variables are exactly analogous, while the variable $v$ takes the place
of time in the heat equation.  We also notice that the source function is
multiplied by a weight factor $-w(v)$\@.

The interpretation of the ``time'' variable $v$ is worth a short discussion.
The variable $v$ increases monotonically with decreasing energy.  Since this
model does not include any reacceleration of positrons, the variable $v$ will
only increase.  This is precisely the situation of the time variable in the
standard heat equation.

In addition to the boundary conditions previously stated, we must also impose
an ``initial'' condition.  It is clear that the initial condition should be
$F(v=0,\vec{x})=0$\@.  This implies that the positron density is zero at $v=0$,
which is infinite energy.

With these things in mind, we can write the solution as an integral over the
source multiplied by a Green's function.  We transform back to the original
energy variable, remembering that Eq.~(\ref{eq:diff-v}) is an equation for
$dn/du$ and not $dn/d\varepsilon$\@.  The solution can be written
\begin{equation}
  \frac{dn}{d\varepsilon}=\taue\varepsilon^{-2}\int
  ^{v(\varepsilon)}_0 dv'\,w(v')\int d^3\vec{x}\,'\,G_{2L}
  \left[v(\varepsilon)-v',\vec{x}-\vec{x}\,'\right]Q(\varepsilon(v'),
  \vec{x}\,')
\end{equation}
where the function $G_{2L}$ is the Green's function for 
Eq.~(\ref{eq:diff-v}) on the slab with thickness $2L$\@.  By our 
construction, $w(v)dv=-d\varepsilon$ and making this variable change 
frees us from the cumbersome Jacobian factor and from implicitly 
inverting the transformation $v(\varepsilon)$\@.  The only place 
$v(\varepsilon)$ remains is in the argument to the Green's function.  
The formal solution is thus written
\begin{equation}
  \frac{dn}{d\varepsilon}=\taue\varepsilon^{-2}\int
  ^\infty_\varepsilon d\varepsilon'\,\int d^3\vec{x}\,'\,G_{2L}
  \left[v(\varepsilon)-v(\varepsilon'),\vec{x}-\vec{x}\,'\right]
  Q(\varepsilon',\vec{x}\,').
  \label{eq:dnde}
\end{equation}

The free-space Green's function for Eq.~(\ref{eq:diff-v}) is given by
\begin{equation}
  G_{\rm free}(v-v',\vec{x}-\vec{x}\,')=
  \left[4\pi K_0\taue(v-v')\right]^{-3/2}
  \exp\left(-\frac{(\vec{x}-\vec{x}\,')^2}
  {4K_0\taue(v-v')}\right)\theta(v-v').
\label{eq:gfree}
\end{equation}
We require a Green's function that vanishes on the planes $z=\pm L$, the
boundaries of our diffusion region.  We can use a set of image charges,
\begin{equation}
  x'_n=x',\quad y'_n=y', \quad z'_n=2Ln+(-1)^nz',
\end{equation}
to find the Green's function,
\begin{equation}
G_{2L}(v-v',\vec{x}-\vec{x}')=\sum_{n=-\infty}^\infty(-1)^nG_{\rm free}
(v-v',\vec{x}-\vec{x}'_n).
\end{equation}
It is simple to verify that this Green's function vanishes on the
boundaries.

Now we choose an appropriate source function $Q$\@.  To simplify the
calculation, we will assume that the source is uniform in $z$ and
depends only on the cylindrical coordinate $r$\@.  Let the number
density of WIMPs in the halo be $n_0g(r_{\rm sph})$, where $r_{\rm
sph}$ is the spherical coordinate and $n_0$ is the number density in
the center of the galaxy.  The annihilation rate, and thus the source
strength, is proportional to the square of the density.  We define
\begin{equation}
  f(r)=\frac{1}{2L}\int^L_{-L}dz\,g^2(r_{\rm sph}).
  \label{eq:feq}
\end{equation}
We will mainly use the isothermal halo profile, 
\begin{equation}
  g_{\rm iso}(r,z)=\frac{a^2}{r_{\rm sph}^2+a^2}=\frac{a^2}{r^2+z^2+a^2},
\end{equation}
with $a=$5 kpc and a local halo density of 0.3 GeV/cm$^3$\@. We
will refer to this is our ``canonical'' halo model.
We will also compare with the Navarro, Frenk and White profile
\cite{nfw},
\begin{equation}
  g_{\rm NFW}(r,z) = \frac{a^3}{r_{\rm sph} \left(r_{\rm sph}+a\right)^2}
\end{equation}
with $a=25$ kpc.
Finally, the source function is
\begin{equation}
  Q(\varepsilon,r)=n_0^2f(r)\sigv\frac{d\phi}{d\varepsilon},
  \label{eq:sourcefcn}
\end{equation}
where $\sigv$ is the annihilation rate and $d\phi/d\varepsilon$ is the
positron spectrum in a single annihilation.  This contains
annihilations both to monochromatic and continuum positrons. Note that
if $\Omega_\chi h^2<0.025$, we need to rescale the source function by
$(\Omega_\chi h^2/0.025)^2$\@. We will only include models with
$\Omega_\chi h^2<0.025$ (with the rescaled source function) in
Fig.~\ref{fig:phiepvsmx} (b).

We now insert this source into the Green's function integral,
Eq.~(\ref{eq:dnde}),
\begin{eqnarray}
  \frac{dn}{d\varepsilon}&=&n_0^2\sigv\taue\varepsilon^{-2}
  \int_\varepsilon^\infty d\varepsilon'\,\left[4\pi K_0\taue(v(\varepsilon)-
  v(\varepsilon'))\right]^{-3/2}\frac{d\phi}{d\varepsilon'}
  \sum_{n=-\infty}^\infty(-1)^n \\
  &&\int_{-L}^Ldz'\,\int_0^\infty dr'\,r'f(r')\int_0^{2\pi}d\theta'\,
  \exp\left(-\frac{r^2+r'^2-2rr'\cos\theta'+
  (z-(-1)^nz'-2Ln)^2}{4K_0\taue(v(\varepsilon)-v(\varepsilon'))}\right).
  \nonumber
\end{eqnarray}
The $z'$ integration will simply yield error functions and the
$\theta'$ integration yields $2\pi I_0$, where $I_{0}$ is the modified Bessel
function of the first kind.
We are left with the $r'$ integral, which we put in the function ${\cal I}$,
\begin{eqnarray}
  {\cal I}[r,z,\deltav] & = & \frac{1}{4K_0\taue\deltav}
  \sum_{n=-\infty}^\infty\sum_\pm\erf\left(\frac{(-1)^nL+2Ln\pm z}
  {\sqrt{4K_0\taue\deltav}}\right)\times \\ \nonumber
  & & \int_0^\infty dr'\,r'f(r')\tilde{I}_0\left(\frac{2rr'}
  {4K_0\taue\deltav}\right)
  \exp\left(-\frac{(r-r')^2}{4K_0\taue\deltav}\right)
\end{eqnarray}
with $\tilde{I}_0(x)=I_0(x)e^{-x}$\@. Note that when $x\gg 1$,
$I_0(x)\approx e^x/\sqrt{2\pi x}$\@.  When $4K_0\taue\deltav \ll L^2$,
this sum and integral is easy to perform, since we have a simple
gaussian and all of the error functions cancel.  The result is
\begin{equation}
  {\cal I}\left[r,z,\deltav\ll L^2/(4K_0\taue)\right]=f(r).
\end{equation}

For clarity, we define an energy dependent diffusion time,
\begin{equation}
  \taud[\deltav] = \taud[\varepsilon,\varepsilon'>\varepsilon]=\taue
  {\cal I}\left[v(\varepsilon)-v(\varepsilon')\right].
\end{equation}
This encodes all of the information about the spatial extent of the source
function and the energy dependence of the diffusion constant.  What this
function represents is the lifetime against energy loss and diffusion of
positrons at a given energy which were emitted at a higher energy.  There are
two regimes to this timescale.  When $\deltav$ is small, the energy lost is
small, thus the particles have not been propagating for long, and have not had
any chance to escape.  At large $\deltav$, much energy has been lost, thus the
particles have been propagating for a long time and have had a chance to escape
the diffusion zone.  In the former regime, it is the energy loss that dominates
the propagation, and in the latter, it is the diffusion.  Diffusion occurs in
both regimes, but in the former, it occurs over a small enough region that the
boundary can be neglected.  At intermediate $\deltav$, there is a slight rise
in the function $\taud$, before it drops precipitously.  This is because the
effective amount of source material increases with increasing $\deltav$\@.  As
more energy is allowed to be lost, the positrons can come from farther away.
This allows us to sample the higher density inner regions of the galaxy.  Only
when the boundary can be probed does $\taud$ begin to drop.

\begin{figure*}[t]
\centerline{\epsfig{file=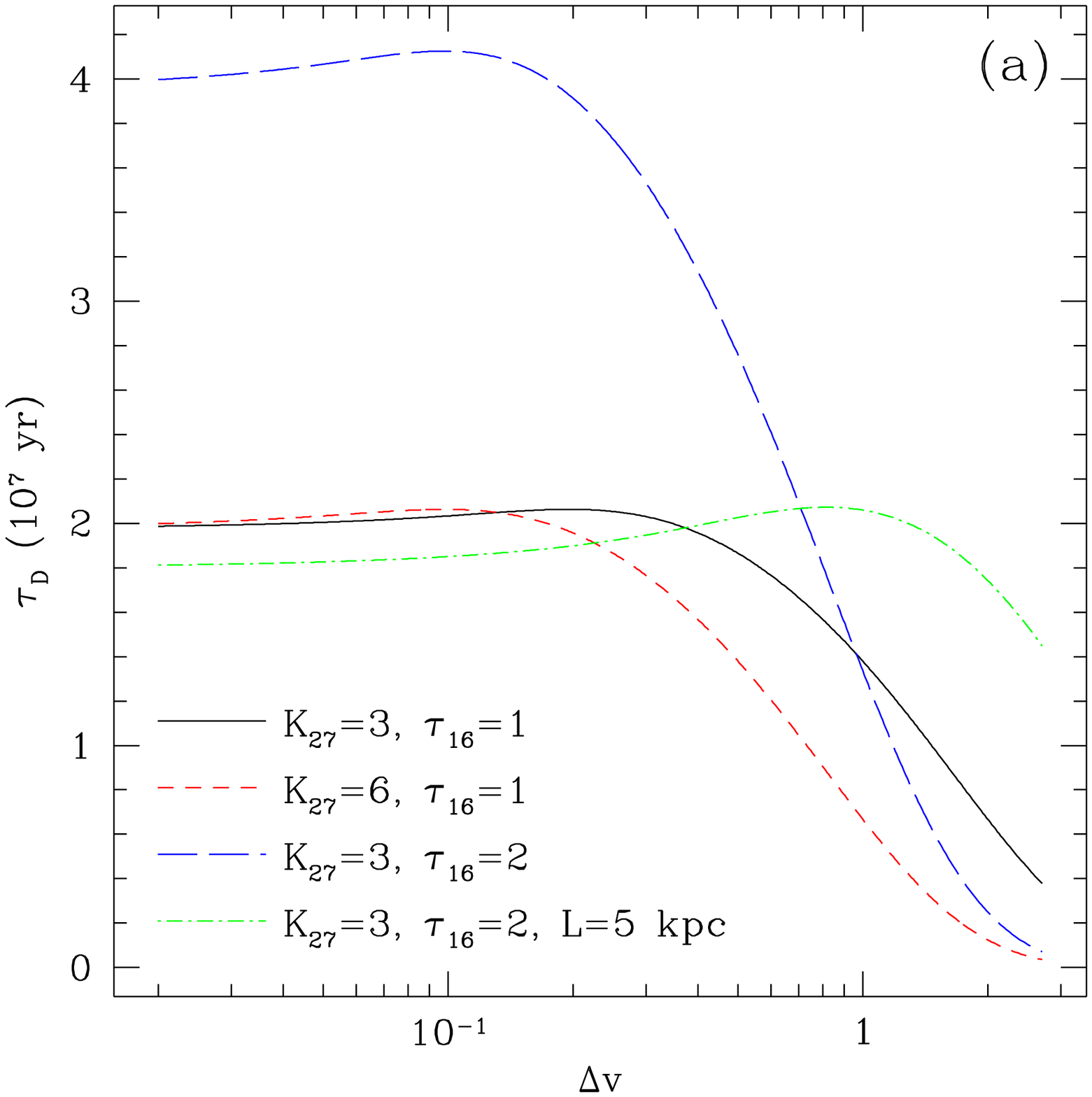,width=0.33\textwidth}
\epsfig{file=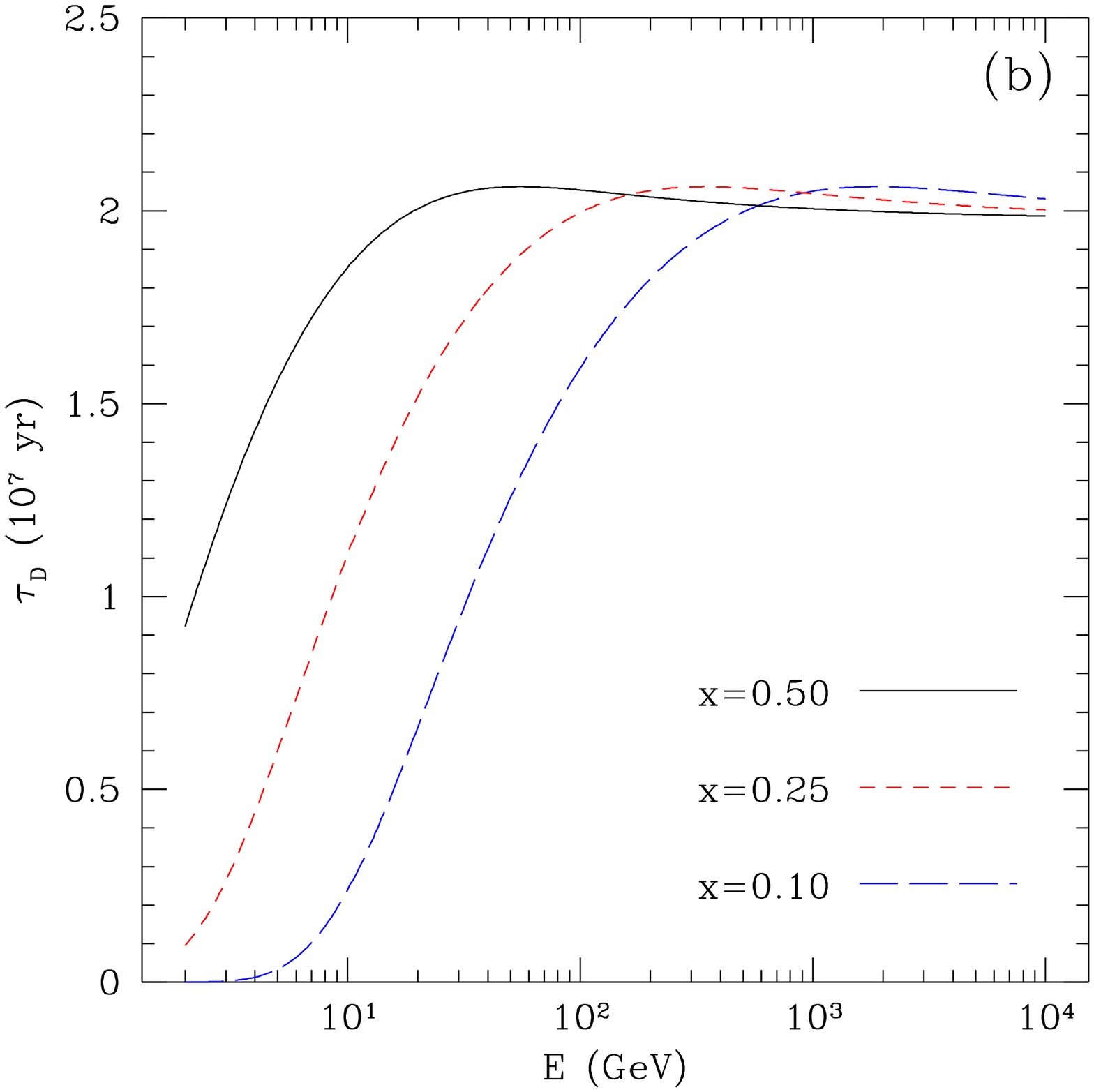,width=0.33\textwidth}
\epsfig{file=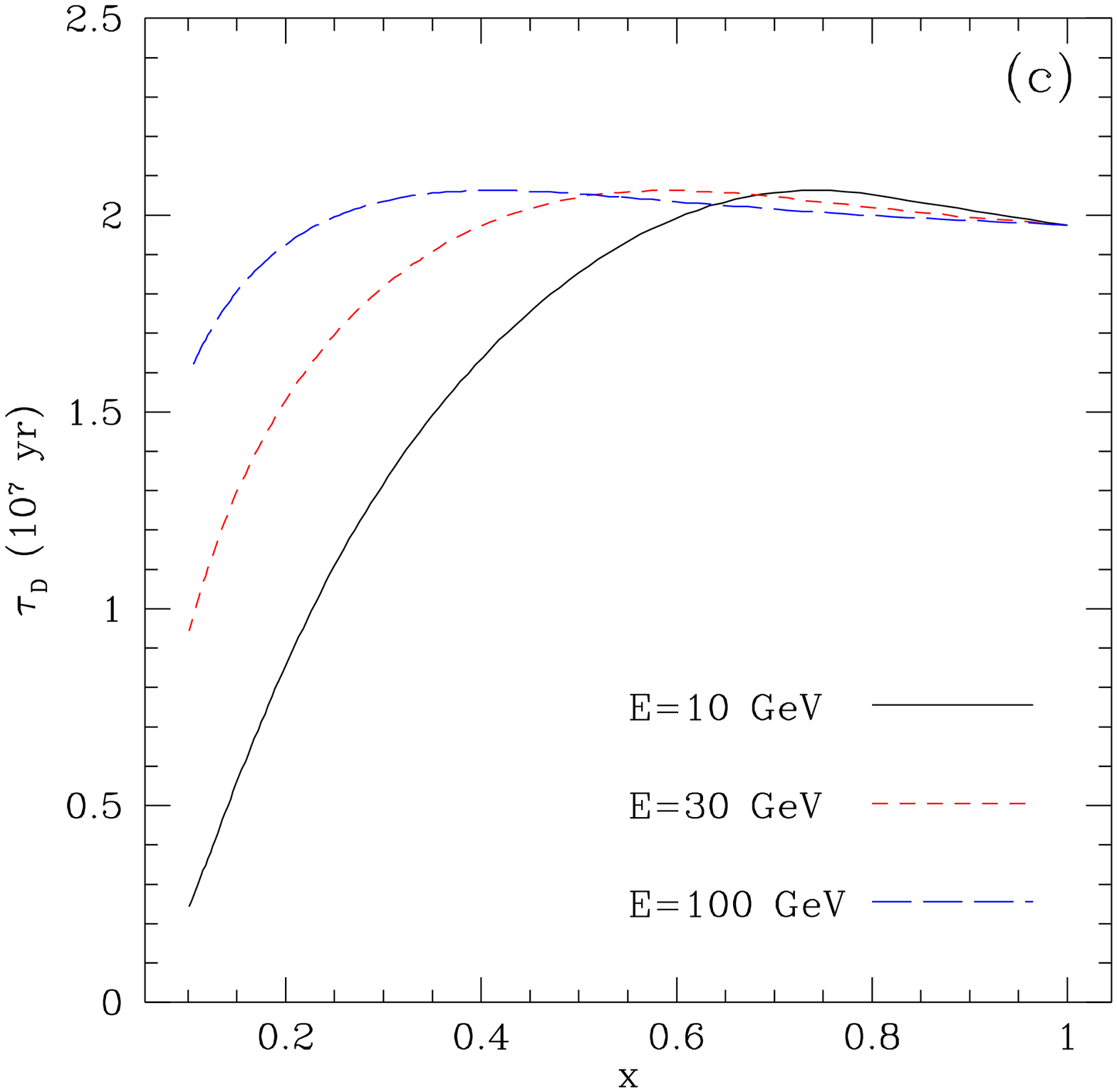,width=0.33\textwidth}} 
\caption{(a) The effective diffusion time $\tau_{D}$ as a function of
$\Delta v$\@.  When doubling the diffusion constant $K_{27}$,
positrons remain trapped for a shorter time before escaping the
galaxy.  At low energy losses (small $\deltav$) this of no importance
since it is the energy loss time that is important in that region.
When doubling the energy loss time $\tau_{16}$ we increase the flux at
low energy losses since the positrons don't lose energy as quickly.
At higher energy losses, the flux goes down since the positrons have
now had more time to escape.  Increasing $L$ to 5 kpc, we see that
diffusion becomes effective at a larger $\deltav$, as expected.  The
positrons require more time to escape the larger diffusion region.
The $10\%$ difference at low $\deltav$ is because the function $f(r)$
depends on $L$\@.  In (b) we show the effective diffusion time
$\tau_{D}(x E,E)$ versus the injection energy $E$ for $x$=0.50, 0.25
and 0.10. In (c) we show the effective diffusion time $\tau_D(x E,E)$
as a function of $x$ for different injection energies. In (b) and (c)
propagation model B is used.}
\label{fig:taud}
\end{figure*}

In Fig.~\ref{fig:taud} (a) we show $\taud$ as a function of $\deltav$
for the position of the solar system, $r=8.5$ kpc and $z=0$ kpc, for
various combinations of $K_{27}=K_0/10^{27}$ cm$^2$ s$^{-1}$,
$\tau_{16}=\taue/10^{16}$ s, and $L$\@.  At low $\deltav$, $\taud$ is
independent of $K_{27}$ and linear in $\taue$, as this is the regime
where energy loss dominates.  At high $\Delta v$, $\taud$ decreases
with increasing $K_{27}$, as expected for the diffusion regime.
Increasing $\taue$ implies that particles have been propagating
longer, and have had more chance to escape.  Increasing $L$ to 5 kpc
increases the signal at high $\deltav$ since it is more difficult to
escape the diffusion zone.  At low $\deltav$, the 10$\%$ discrepancy
is due to the fact that $f(r)$ depends on $L$\@.  The small size of
the discrepancy indicates that using a source function uniform in $z$
is adequate.  In Fig.~\ref{fig:taud} (b) we show the function
$\tau_{D}(x E,E)$ versus the injection energy $E$ for $x$=0.50, 0.25
and 0.10.  In Fig.~\ref{fig:taud} (c) we show $\tau_D(x E,E)$ versus
$x$ for some different injection energies $E$\@. In both
Fig.~\ref{fig:taud} (b) and (c), propagation model B is used.

The positron spectrum is now given by
\begin{equation}
  \frac{dn}{d\varepsilon}=n_0^2\sigv\varepsilon^{-2}
  \int_\varepsilon^\infty d\varepsilon'\,\frac{d\phi}{d\varepsilon'}
  \taud\left[\varepsilon,\varepsilon'\right],
\end{equation}
giving the total positron spectrum
\begin{equation}
  \frac{dn}{d\varepsilon}=n_0^2\sigv\varepsilon^{-2}\Bigg\{B_{\rm line}\taud
  \left[\varepsilon,\mct\right]+\int_{\varepsilon}^{\mct}d\varepsilon'\,
  \left.\frac{d\phi}{d\varepsilon'}\right|_{\rm
  cont.}\taud[\varepsilon,\varepsilon']
  \Bigg\}\theta\left(\mct-\varepsilon\right).
\label{eq:dndefinal}
\end{equation}
Remembering that this is an expression for the number density of positrons, the
flux is given by
\begin{equation}
  \frac{d\Phi}{d\varepsilon} = \frac{\beta c}{4 \pi} \frac{dn}{d\varepsilon}
  \simeq \frac{c}{4 \pi} \frac{dn}{d\varepsilon},
\end{equation}
where $\beta c$ is the velocity of a positron of energy 
$\varepsilon$\@.  For the energies we are interested in, $\beta c 
\simeq c$ is a very well justified approximation.

We find that it is numerically advantageous to perform the integral in the
variable $v$\@.  Replacing the Jacobian factor, we find
\begin{equation}
  \frac{dn}{d\varepsilon}=n_0^2\sigv\varepsilon^{-2}\Bigg\{B_{\rm line}
  \taud\left[\varepsilon,\mct\right]+\int_{v(\mct)}^{v(\varepsilon)}dv'\,
  w(v')\left.\frac{d\phi}{d\varepsilon}\right|_{\rm cont.}[\varepsilon(v')]
  \taud[\varepsilon,\varepsilon(v')]\Bigg\}\theta\left(\mct-\varepsilon\right).
\end{equation}
The function $\taud = \taue {\cal I}$ can be tabulated once.  The
integrand is smooth, and is thus easily computed for a range of
values, equally spaced in $\ln\deltav$\@.  Likewise the sum on error
functions converges rapidly, and for the range of $\deltav$ values we
are concerned with, need not be taken past $n=\pm 10$\@.

\subsection{Solar modulation}

The formalism developed above gives the interstellar flux of positrons from
neutralino annihilations.  There is a further complication in that interactions
with the solar wind and magnetosphere alter the spectrum.  These effects are
referred to as solar modulation.  This can be neglected at high energies, but
at energies below about 10 GeV, the effects of solar modulation become
important. We will not consider solar modulation, since its effects can be
removed by considering the positron fraction, $e^+/(e^+ + e^-)$, instead of the
absolute positron fluxes.

However, another complication is the possibility that the solar
modulation is charge-sign dependent as indicated in Ref.\ \cite{clem}.
Following their treatment we can write the positron fraction at Earth
in a solar cycle $A^{+}$ and $A^{-}$ respectively as
\begin{eqnarray}
  f^+_E&=&\frac{f^2(R+1)-f}{R(2f-1)}, \label{eq:feplus} \\
  f^-_E&=&\frac{f^2(R+1)-fR}{2f-1}. \label{eq:feminus}
\end{eqnarray}
where $f$ is the interstellar positron fraction and $R$ is the ratio of 
the total flux of positrons and electrons in a solar $A^+$ cycle to 
that in an $A^-$ cycle.  The value $R$ can be measured, and is 
approximately given by \cite{clem}
\begin{equation}
  R(\varepsilon)=\max 
  \left[\min\left(0.45+0.17\;\ln\varepsilon,1\right),0.18 \right].
\end{equation}

We notice that in the absence of charge-sign dependence, the positron
fraction is unaffected by solar modulation effects. As we will see
later, the charge-sign dependent solar modulation worsens the
agreement between the measurements and the expected background and
given the large uncertainties involved, we will not use these
expressions when predicting the positron fraction from neutralino
annihilation.

\begin{figure}[t]
\centerline{\epsfig{file=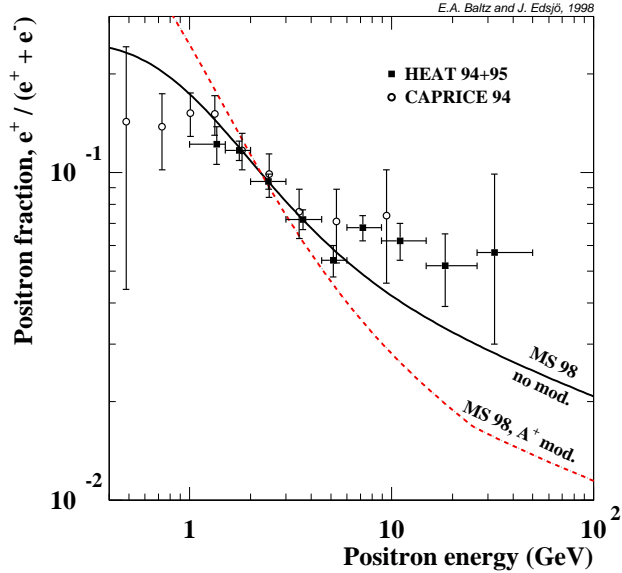,width=0.49\textwidth}}
\caption{The positron fraction for the background is shown both
without and with charge-sign dependent solar modulation
\protect\cite{clem}.  The normalization has been kept free and is
fitted to the {\sc Heat} 94+95 data \protect\cite{heatfrac}. Also
shown is the {\sc Caprice} 94 data \protect\cite{caprice}. The
backgrounds are parameterizations to the results by MS
\protect\cite{MoskStrong98}.}
\label{fig:epbkg}
\end{figure}

\subsection{Positron and electron backgrounds}

When we want to compare our predictions with experiments, we must
consider the expected background of positrons.  Since most experiments
measure the positron fraction and not the absolute positron flux, we
also need to consider the electron background.  We will use the most
recent calculation by Moskalenko and Strong \cite{MoskStrong98}
(hereafter MS)\@.  For their model 08-005 without reacceleration, we
have parameterized their calculated primary electron and secondary
electron and positron fluxes as follows:
\begin{eqnarray}
  \left(\frac{d\Phi}{dE}\right)_{{\rm prim.~} e^{-}~\rm bkg} & = &
  \frac{0.16 \,\varepsilon^{-1.1}}
  {1+11 \,\varepsilon^{0.9}+3.2 \,\varepsilon^{2.15}}
  \; {\rm GeV}^{-1}\;{\rm cm}^{-2}\;{\rm s}^{-1}\;{\rm sr}^{-1},\\
  \left(\frac{d\Phi}{dE}\right)_{{\rm sec.~} e^{-}~\rm bkg} & = &
  \frac{0.70 \,\varepsilon^{0.7}}
  {1+110 \,\varepsilon^{1.5}+600\,\varepsilon^{2.9}+580 \,\varepsilon^{4.2}}
  \; {\rm GeV}^{-1}\;{\rm cm}^{-2}\;{\rm s}^{-1}\;{\rm sr}^{-1},\\
  \left(\frac{d\Phi}{dE}\right)_{{\rm sec.~} e^{+}~\rm bkg} & = &
  \frac{4.5 \,\varepsilon^{0.7}}
  {1+650 \,\varepsilon^{2.3}+1500 \,\varepsilon^{4.2}}
  \; {\rm GeV}^{-1}\;{\rm cm}^{-2}\;{\rm s}^{-1}\;{\rm sr}^{-1}
  \label{eq:epsec}
\end{eqnarray}
where we as before have used $\varepsilon=E/(1~{\rm GeV})$\@.  These
parameterizations agree with their curves to within 10--15\% for the
whole intervals given in Ref.~\cite{MoskStrong98} (approximately
0.001--1000 GeV for primary electrons and 0.01--100 GeV for the
secondary electrons and positrons).  The parameterizations also have
the same asymptotic slopes as the calculated fluxes at both the low
and high energy end and extrapolations should be acceptable if not too
far out of the regions given above. These predictions also agree
roughly with the absolute flux measurements by the {\sc Heat}
experiment \cite{heatabs}. The experimental error bars are however
smaller on the measurements of the positron fraction and in
Fig.~\ref{fig:epbkg} we have made a fit to the {\sc Heat} 94+95
measurements \cite{heatfrac} of the positron fraction from the
background alone keeping only the normalization of the positron flux
free.  We also show the best fits with charge-dependent solar
modulation included.  The normalization factors for the best fits are
$k_b = 1.11$ and $k_b = 0.609$ without charge-dependent solar
modulation and with $A^+$ modulation respectively ($A^{+}$ is the
correct state for the {\sc Heat} measurement).  The corresponding
reduced $\chi^2$ are 3.14 and 10.8, i.e.\ not very good fits. We also
show the {\sc Caprice} 94 measurements of the positron fraction
\cite{caprice}.

We clearly see that the background is too steep at low energies and
including the charge-sign dependent solar modulation only worsens the
fit.  For models with reacceleration, the fit is worse
still\cite{MoskStrong98}.  We also note that there is an indication in
the {\sc Heat} data of an excess at 6--50 GeV\@.  However, one should
keep in mind that (a) the uncertainties below 10 GeV are large due to
our poor knowledge of the solar modulation effects, (b) it is possible
to make a better fit by hardening the interstellar nucleon spectrum
\cite{MoskStrong98} (even though this might be in conflict with
antiproton measurements above 3 GeV\cite{MSRdiffgamma}) and (c) it is
possible to make a better fit by hardening the primary electron
spectrum (even though this might be in conflict with direct electron
measurements at higher energies \cite{MoskStrong98})\@.  We will not
consider these issues here.  Instead we will assume the background as
given by the parameterizations above without charge-dependent solar
modulation.  We will then compare the positron background with our
predictions and we also make a simultaneous fit of the positron
fraction background signal from neutralino annihilations having only
the normalization of the background and signal positron fluxes as free
parameters.

To conclude the discussion of the backgrounds, there is clearly room
(or even an indication) of an excess of positrons at intermediate
energies that might be due to neutralino annihilations. However, the
uncertainties are large and other explanations are feasible.  We will
entertain the possibility that the excess is due to neutralino
annihilation in the next section.


\begin{figure*}[t]
\centerline{\epsfig{file=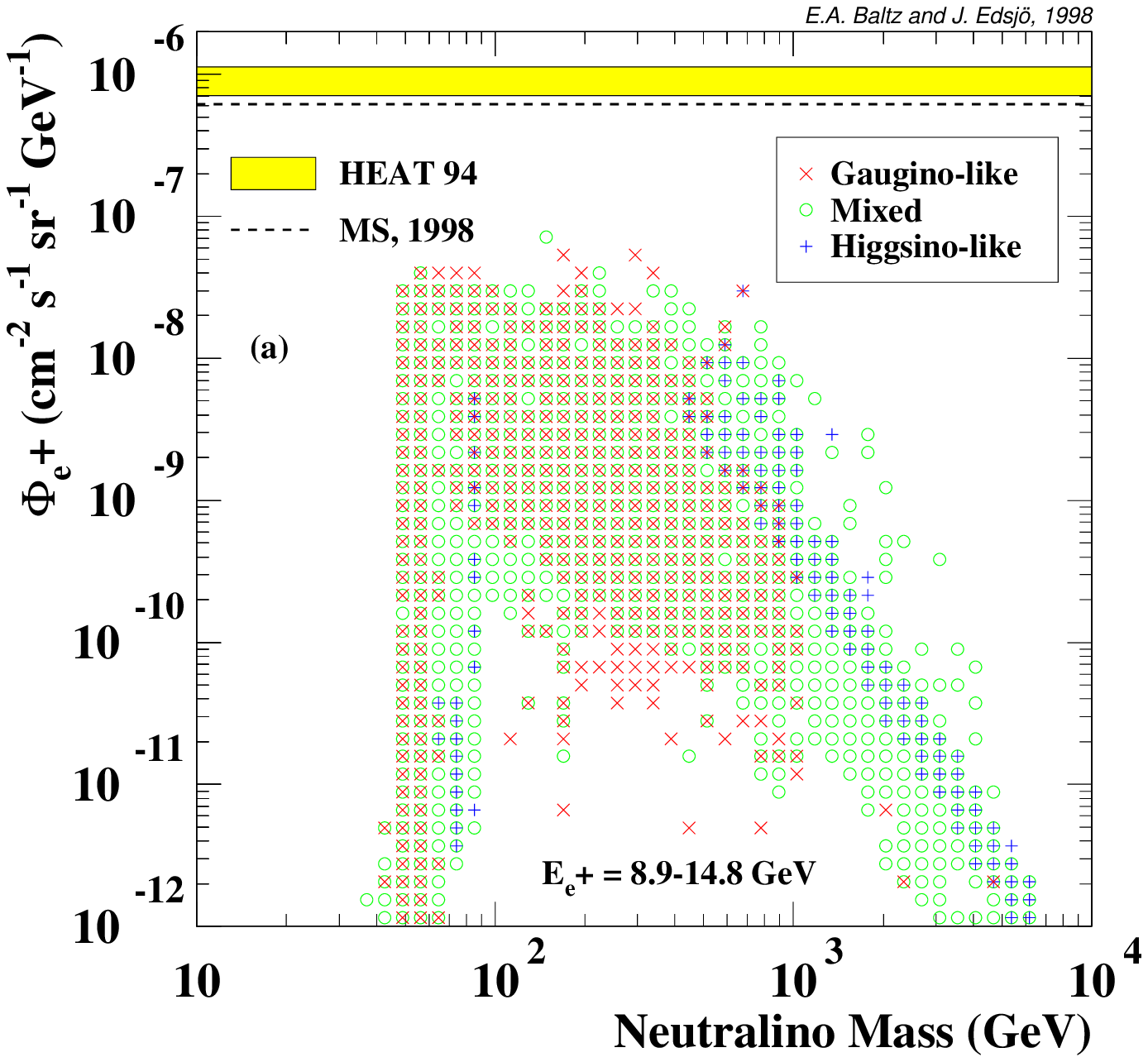,width=0.49\textwidth}
\epsfig{file=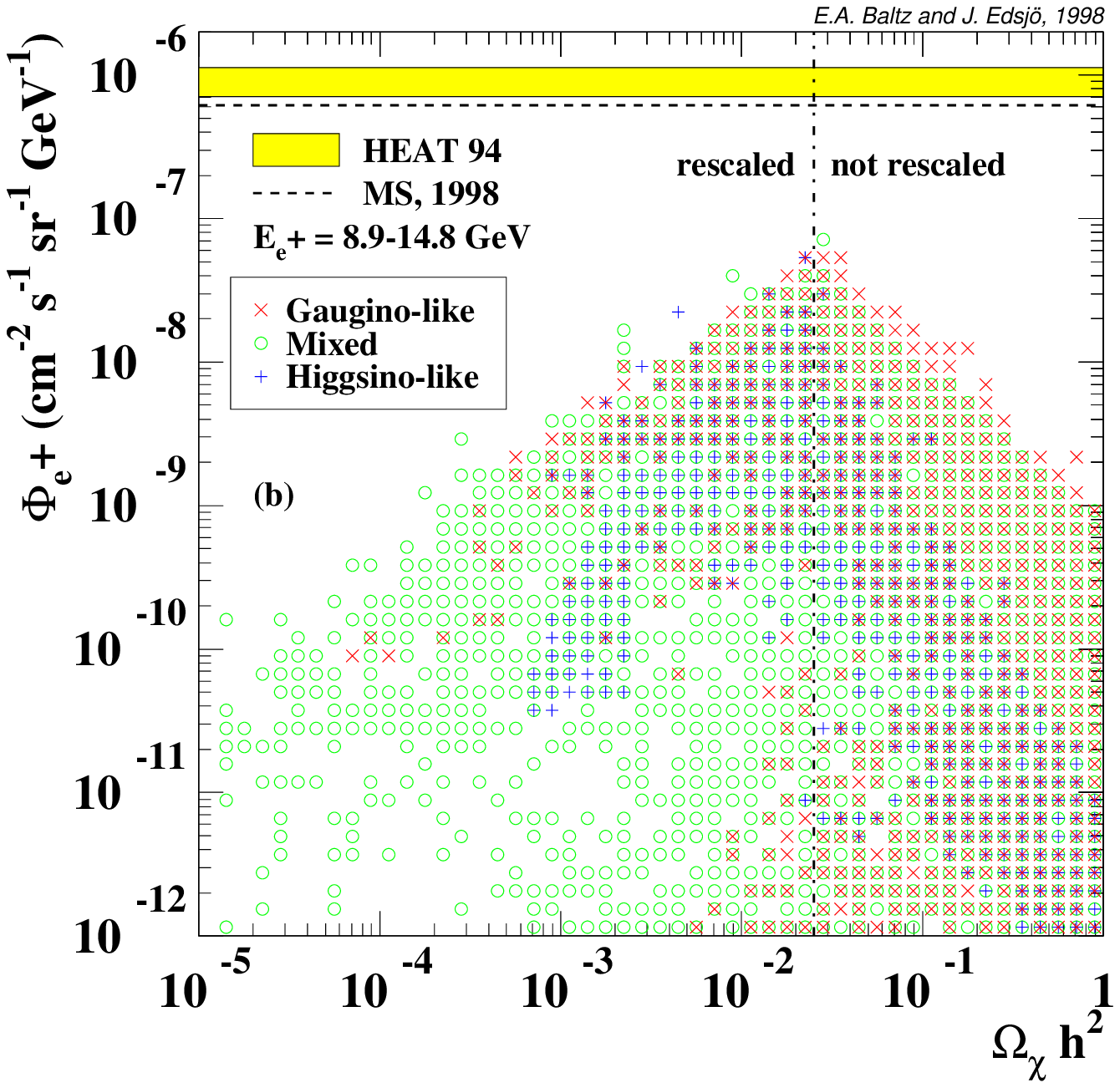,width=0.49\textwidth}} 
\caption{The flux of positrons from neutralino annihilation in the
halo versus (a) the neutralino mass and (b) the neutralino relic
density $\Omega_{\chi}h^{2}$\@.  In (a) only models with
$0.025<\Omega_\chi h^2 < 1$ have been included and in (b) models with
$\Omega_\chi h^2<0.025$ have been rescaled by $(\Omega_\chi
h^2/0.025)^2$\@.  The {\sc Heat} 94 data in this energy interval is
shown together with the background prediction by MS
\protect\cite{MoskStrong98} (as given by our parameterization,
Eq.~(\protect\ref{eq:epsec})).}
\label{fig:phiepvsmx}
\end{figure*}

\begin{figure}[t]
\centerline{\epsfig{file=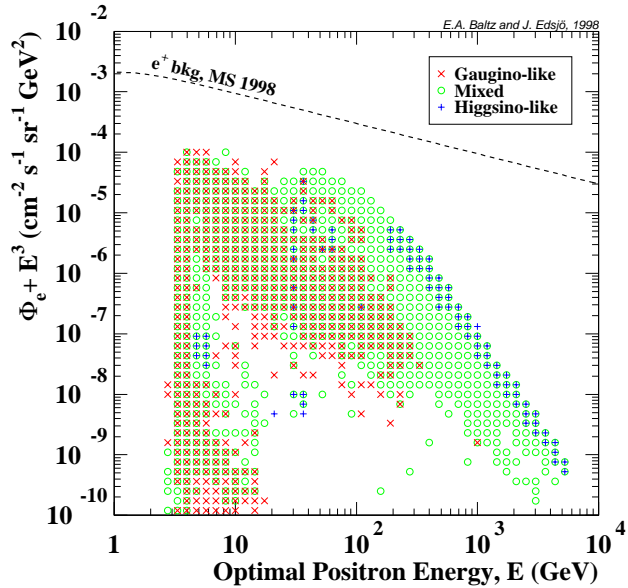,width=0.49\textwidth}}
\caption{The flux of positrons, $\Phi_{e^{+}}E^{3}$ versus the optimal
positron energy, $E$, for which the ratio of the signal flux to the
background flux is highest.  The positron background from MS
\protect\cite{MoskStrong98} (as given by our parameterization,
Eq.~(\protect\ref{eq:epsec})) is shown as the dashed line. Above 100
GeV, the background is an extrapolation of the results by MS.}
\label{fig:phiepopt}
\end{figure}

\begin{figure*}[t]
\centerline{\epsfig{file=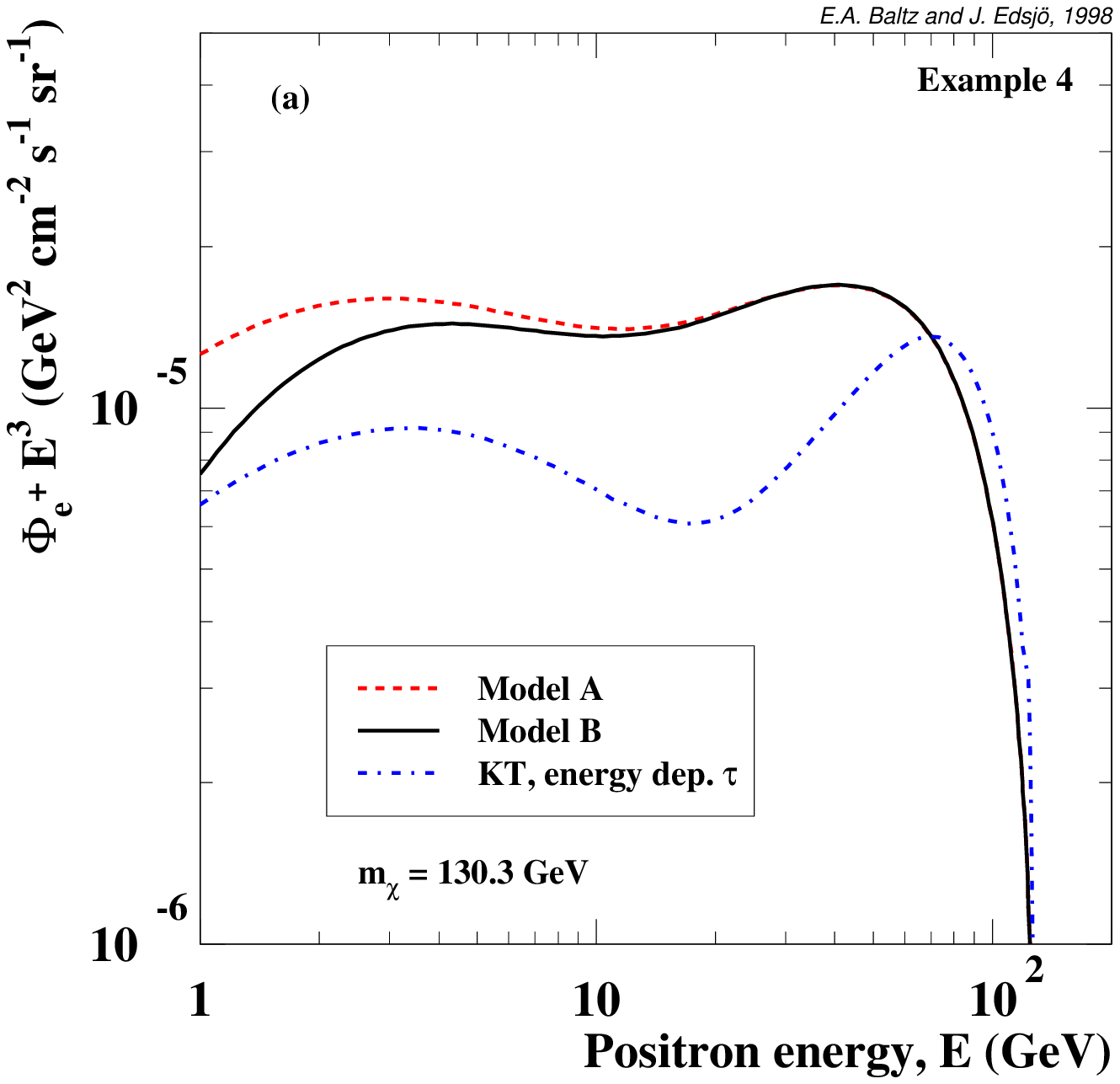,width=0.33\textwidth}
\epsfig{file=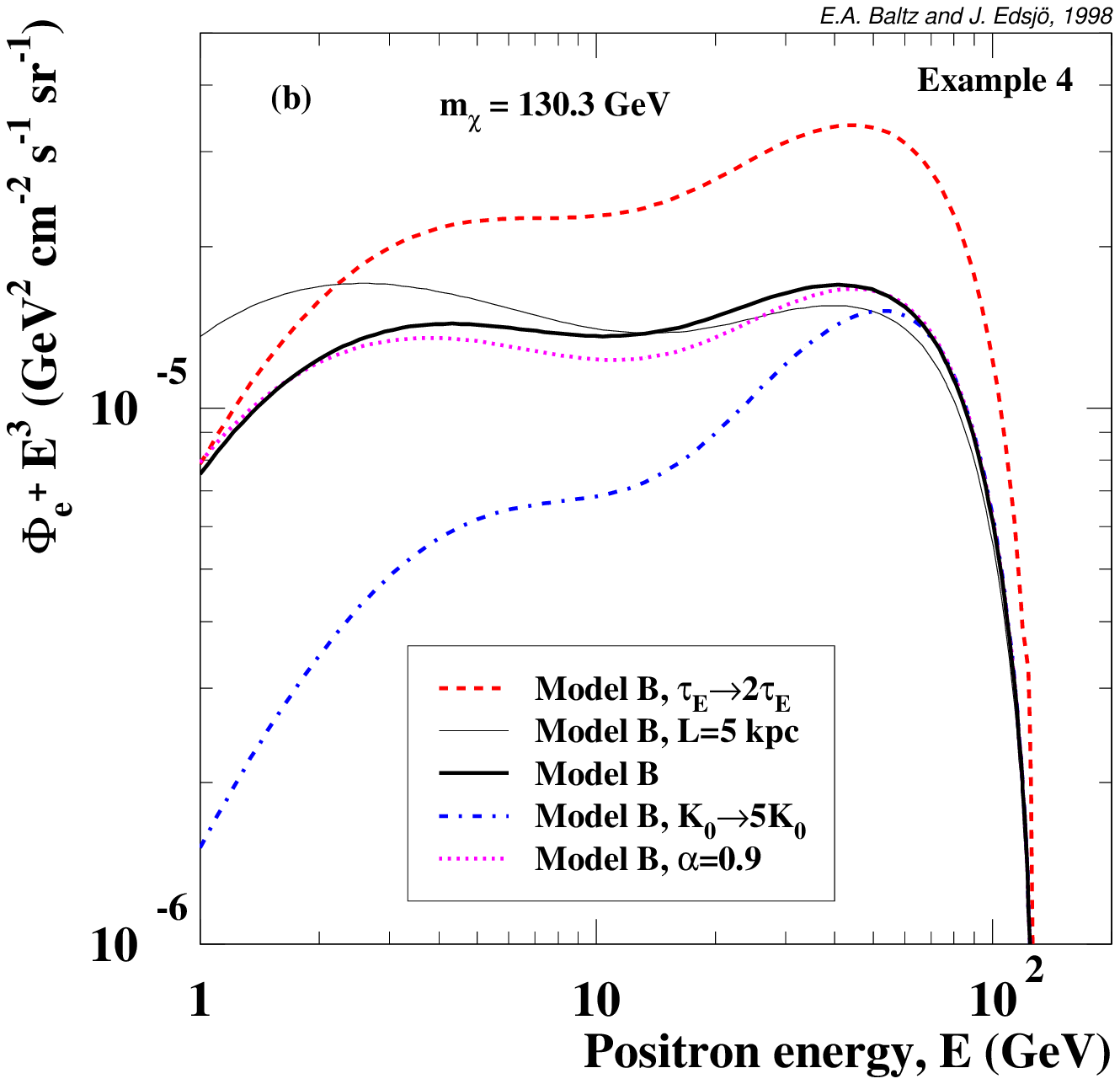,width=0.33\textwidth}
\epsfig{file=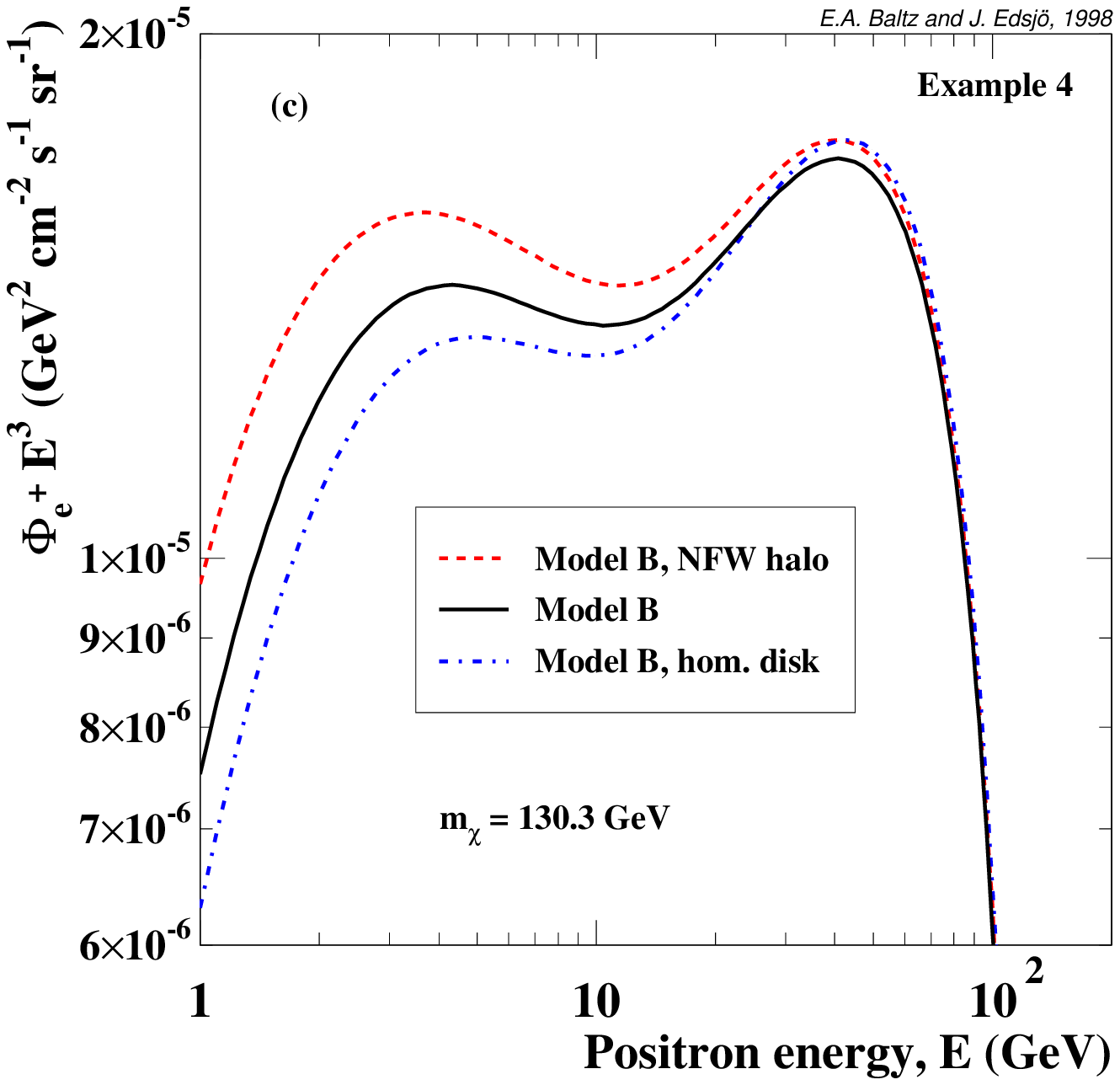,width=0.33\textwidth}}
\caption{An example of a light model annihilating mainly into
$W^{+}W^{-}$\@. In (a), the differences between our models A and B and
the KT model with energy dependent escape time
\protect\cite{kamturner} are shown. In (b) the effects (for model B)
of increasing $K$, $\tau_{E}$, $\alpha$ and $L$ are given and in (c),
the differences between different halo profiles are shown. Note that
the scale is different in (c).}
\label{fig:epcfa}
\end{figure*}

\begin{table}
  \begin{tabular}{clrrrrrrrrll} 
  & & \multicolumn{7}{c}{Parameters and units} &
  \multicolumn{3}{c}{$\chi$ properties} \\ \cline{3-9} \cline{10-12}
  \multicolumn{1}{l}{Example} & Model & 
  \multicolumn{1}{c}{$\mu$} & \multicolumn{1}{c}{$M_{2}$} &
  \multicolumn{1}{c}{$\tan \beta$} & \multicolumn{1}{c}{$m_{A}$} & 
  \multicolumn{1}{c}{$m_{0}$} & \multicolumn{1}{c}{$A_{b}/m_{0}$} & 
  \multicolumn{1}{c}{$A_{t}/m_{0}$} &
  \multicolumn{1}{c}{$m_\chi$} & \multicolumn{1}{c}{$Z_g$} &  
  \multicolumn{1}{c}{\raisebox{0ex}[3ex][0ex]{$\Omega_{\chi}h^{2}$}} \\
  \multicolumn{1}{l}{number} & number & 
  \multicolumn{1}{c}{GeV} & \multicolumn{1}{c}{GeV} & 
  \multicolumn{1}{c}{1} & \multicolumn{1}{c}{GeV} & \multicolumn{1}{c}{GeV} & 
  \multicolumn{1}{c}{1} & \multicolumn{1}{c}{1} & \multicolumn{1}{c}{GeV} & 
  \multicolumn{1}{c}{1} & \multicolumn{1}{c}{1} \\ \hline 
  1 &  JE28\_\_005683  & $-852.3$ & $-670.1$ & 13.1 & 664.0 & 1940.6
  & $-1.56$ & $-1.60$ & 335.7 & 0.9951 & 0.029 \\
  2 &  JEsp3\_000426   & 2319.9 & 4969.9 & 40.6 & 575.1 & 2806.6
  & 2.0 & 0.27 & 2313.3 & 0.028 & 0.16 \\
  3 &  JE29\_\_034479  & $-1644.3$ & $-202.2$ & 54.1 & 181.7 &
  2830.9 & 2.57 & 2.50 & 101.6 & 0.99927 & 0.042 \\
  4 &  JE23\_\_000159  & $-221.8$ & $-324.5$ & 1.01 & 792.2 &
  2998.7 & 1.71 & 0.67 & 130.3 & 0.660 & 0.058 \\
  \end{tabular} 
\caption{Example models.  The parameter values are given together with
the neutralino masses, gaugino fractions and relic densities.}
\label{tab:examples}
\end{table}

\begin{figure*}[t]
\centerline{\epsfig{file=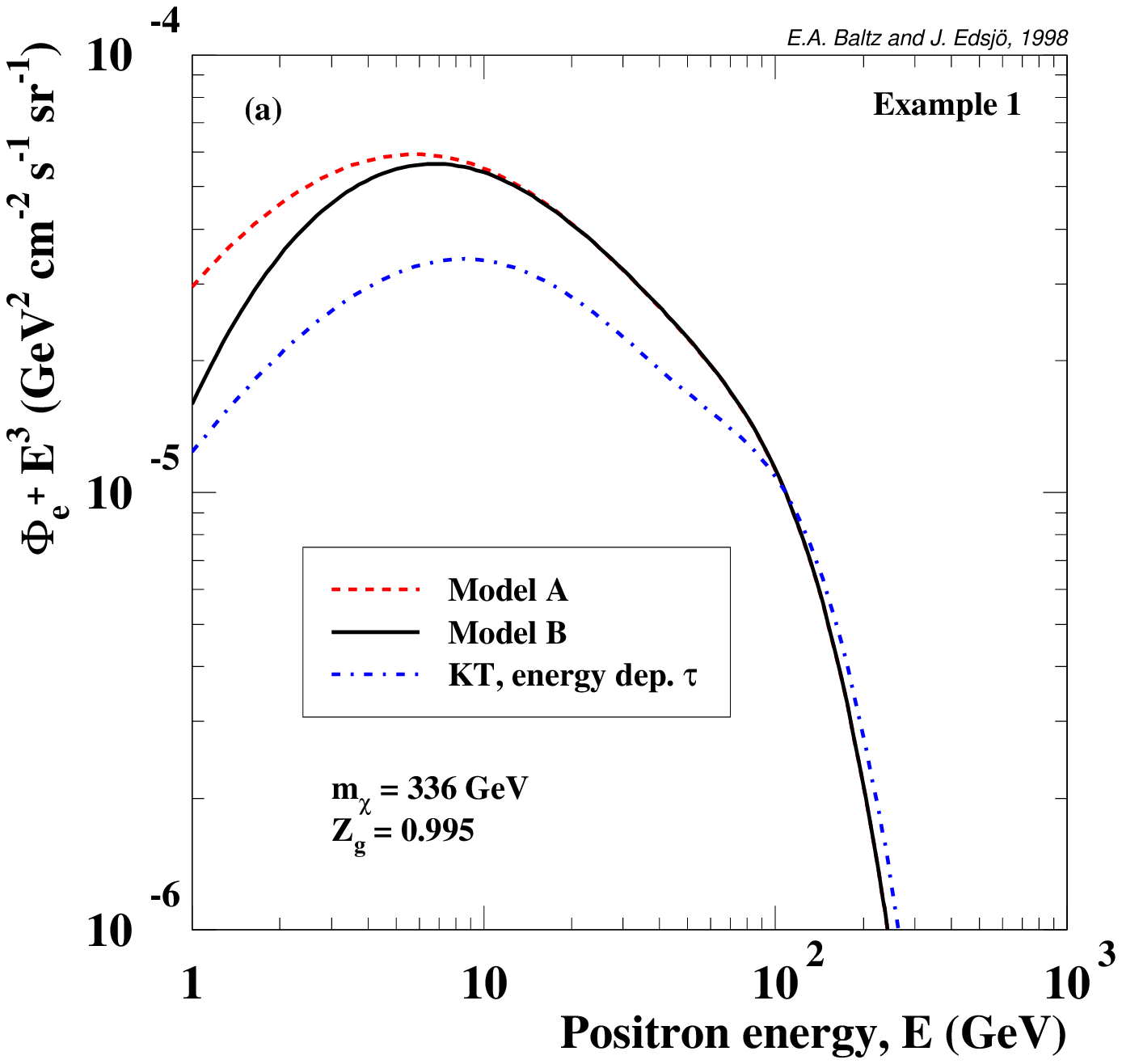,width=0.33\textwidth}
\epsfig{file=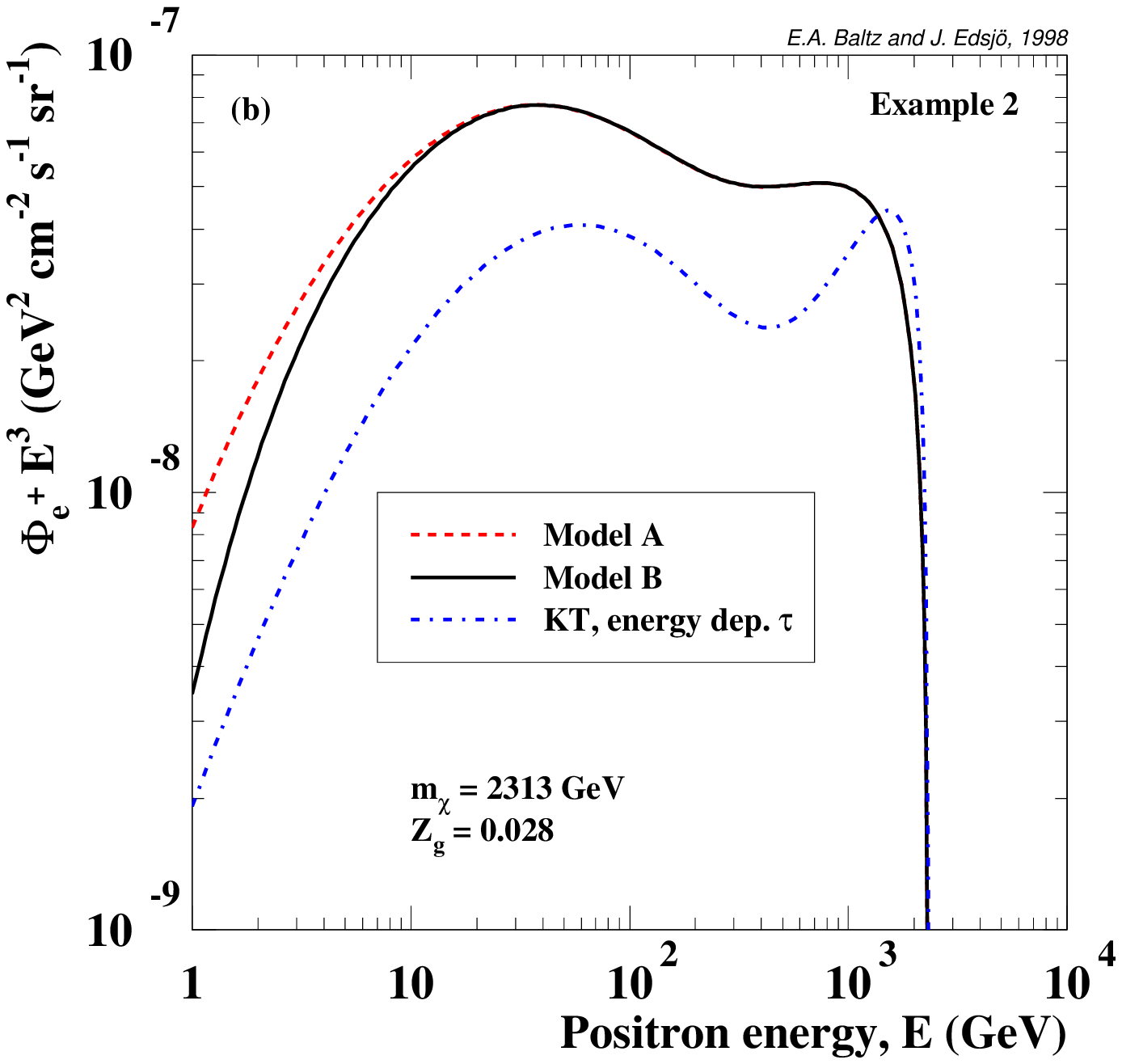,width=0.33\textwidth}
\epsfig{file=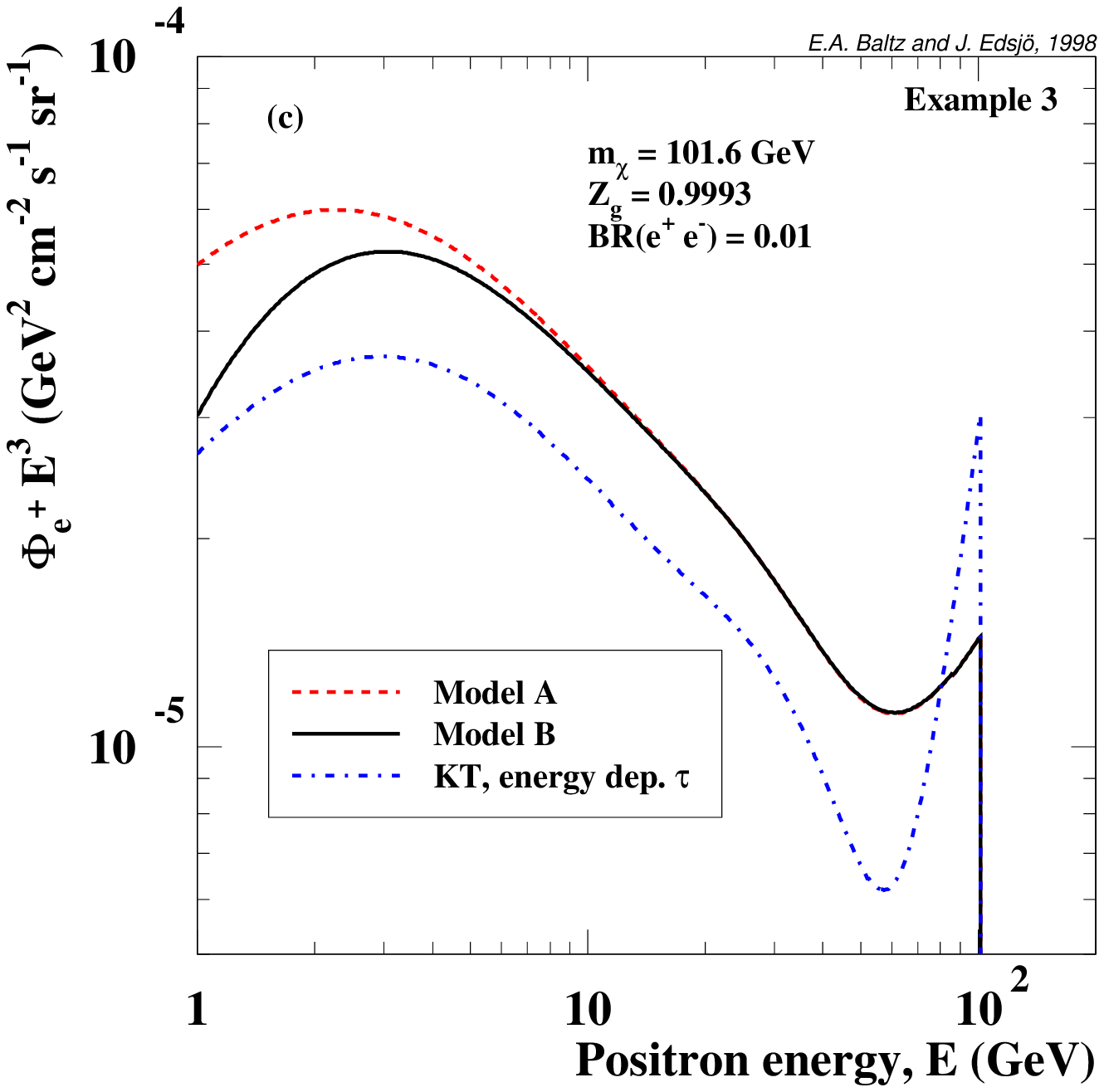,width=0.33\textwidth}}
\caption{Example of positron spectra versus the positron energy.  In
(a) we show a typical spectrum for medium-heavy neutralinos, in (b) we
show a spectrum from a heavy neutralino when gauge bosons dominate the
annihilation final states and in (c) a spectrum when the branching
ratio for annihilation into $e^{+}e^{-}$ has been increased to 0.01.
In all spectra we show them as calculated with our two models and with
the energy-dependent model by KT \protect\cite{kamturner}}
\label{fig:epcfb}
\end{figure*}

\section{Results}

We are now ready the present our results and compare with the expected
background and experimental measurements. We will first consider the
absolute fluxes and investigate the dependence on the propagation
parameters. We will use our propagation model B with the isothermal
sphere as our default model. We will also show some typical
spectra. In the next subsection we will then discuss the predicted
positron fractions and compare with the experiments, in particular the
excess at 6--50 GeV indicated by {\sc Heat} data \cite{heatfrac}.

\subsection{Absolute fluxes and spectra}

As a representative energy we will choose the average flux in the
energy range 8.9--14.8 GeV which corresponds to one {\sc Heat} bin
(with average energy of about 11 GeV).  In this energy range, we
should be fairly unaffected by solar modulation effects.  In
Fig.~\ref{fig:phiepvsmx} we show the absolute positron flux for
propagation model B versus the neutralino mass and versus the relic
density $\Omega_{\chi} h^{2}$\@.  We also compare with the {\sc Heat}
1994 measurement of the absolute flux \cite{heatabs}.  We clearly see
that with our canonical halo model we find no models that give fluxes
as high as those measured.  However, one should keep in mind that we
probably have an overall uncertainty of about an order of magnitude
coming from uncertainties regarding the propagation.  We also have
uncertainties coming from our lack of knowledge of the halo structure.
For example, the halo could be clumpy which could easily increase the
signal by orders of magnitude \cite{silkstebbins,clumpy}.  The local
halo density is also uncertain by at least a factor of 2 (which
corresponds to a factor of 4 uncertainty in the positron flux).  In
Fig.~\ref{fig:phiepvsmx} (a) we see that once we are above the $W$
mass, the spread of the predictions is much smaller.  We also see that
for heavy neutralinos, Higgsinos and mixed neutralinos typically give
higher fluxes than the gaugino-like neutralinos.  This is because the
annihilation cross section to gauge bosons is very suppressed for
gaugino-like neutralinos.  For Higgsinos, the cross section to gauge
bosons is also suppressed, but usually dominates anyway.

In Fig.~\ref{fig:phiepvsmx} (b) we see that the highest fluxes are
approximately proportional to $1/\Omega_\chi h^2$\@.  This is because
$\Omega_\chi h^2$ is approximately proportional to $\langle \sigma v
\rangle^{-1}$ where $\langle \sigma v \rangle$ is the thermally
averaged annihilation cross section.  The highest fluxes for a given
$\Omega_\chi h^2$ are obtained when the annihilation proceeds mainly
to channels giving high positron fluxes and the velocity dependence of
the cross section is small (i.e.\ the thermally averaged cross section
is close to the annihilation cross section at rest).  In that case, it
is essentially the same cross section that determines both
$\Omega_\chi h^2$ and the positron fluxes and hence the strong
(anti)correlation.  When $\Omega_\chi h^2 < 0.025$ we rescale the flux
by $(\Omega_\chi h^2/0.025)^2$ and hence the fluxes for low
$\Omega_\chi h^2$ are proportional to $\Omega_{\chi}h^{2}$\@.  Note
that in all other figures, we will only show the models where
$0.025<\Omega_\chi h^2 <1$\@.

In Fig.~\ref{fig:phiepopt} we show the positron flux versus the
optimal positron energy, $E_{e^+}^{\rm opt}$, where we define
$E_{e^+}^{\rm opt}$ as the energy at which $(d\Phi/dE)_{\rm
signal}/(d\Phi/dE)_{\rm background}$ is maximal.  We also show the
positron background prediction by MS \cite{MoskStrong98}.  We see that
in many cases it is advantageous to look at higher energies than
present experiments do.  Again, we see that for heavier neutralinos,
the mixed and Higgsino-like neutralinos give the highest fluxes.

In Fig.~\ref{fig:epcfa} we show an example of a spectrum when the
neutralino mainly annihilates into $W^{+}W^{-}$\@.  This model is
example 4 in Table~\ref{tab:examples}\@.  We choose this model as an
example to explore the dependence of parameters and propagation models
since it has two nice bumps from both primary as well as secondary and
higher decays/hadronizations of the $W$ bosons.  In (a) we compare our
model A and B with the propagation model of KT \cite{kamturner} with
energy dependent escape time.  Compared to our model B, the KT
propagation model assumes (i) that the diffusion constant coefficient
$\alpha=1$, (ii) that there is no low energy cutoff of the diffusion
constant (i.e.\ as our model A), (iii) that the energy-loss time
$\tau_{E}$ is higher, (iv) that the diffusion constant $K_{0}$ is
higher and (v) that the disk is homogeneous. They also use a leaky-box
model instead of a diffusion model.

In Fig.~\ref{fig:epcfa} (a), we see that our model A gives higher 
fluxes than model B below about 10 GeV\@.  This is easy to understand 
since model A has a lower diffusion constant at low energies and hence 
the positrons stay around longer and we get higher fluxes.  On the 
other hand, the KT propagation gives a sharper peak from the primary 
decay of $W^{+}$ to $e^{+}$\@.  This is, as we will see in figure (b) 
and (c), due to the different values of $K_{0}$, $\tau_{E}$ and 
$\alpha$\@.  At low energies, the KT model has about the same shape as 
our model A, since neither $\tau_{E}$ nor the exact value of $\alpha$ 
are of big importance there.  The different normalizations at low 
energies are due to different values of $K_{0}$\@.

In Fig.~\ref{fig:epcfa} (b) we show what happens in model B if we
change the value of $K_{0}$, $\tau_{E}$, $\alpha$ or $L$\@.
Increasing $K_{0}$ reduces the flux at lower energies (where diffusion
is important), whereas increasing $\tau_{E}$ increases the flux at
higher energies (where energy-loss is more important than escape
through diffusion).  We actually saw this behavior already in
Fig.~\ref{fig:taud} (a) where $\taud$ was shown versus $\deltav$ for
different values of $K_{0}$ and $\tau$\@.  Increasing $\alpha$
essentially tilts the spectrum clockwise at lower energies where
diffusion is important.  Hence, the dip at intermediate energies is
more pronounced.  Increasing $L$ has more or less the same effect as
lowering $K_0$, except at higher energies, where we see a decrease in
the flux.  This is merely an artifact of the averaging over $z$ in
Eq.~(\ref{eq:feq}).  When $L$ increases, the average at a given $r$
decreases. We saw this behavior earlier in Fig.\ \ref{fig:taud} (a).

In Fig.~\ref{fig:epcfa} (c) we show the dependence on the actual halo 
profile.  We expect to mainly see the positrons that are created 
within a few kpc of us, but as we go lower in energy (and hence higher 
in energy loss), the visible region expands.  In (c) we show the 
difference between the isothermal sphere and a homogeneous disk.  The 
isothermal sphere gives higher fluxes at low energies, i.e.\ we see 
the galactic core at lower energies.  If we use a steeper halo profile 
like the Navarro, Frenk and White profile \cite{nfw}, the flux at 
lower energies goes up even more.  The dependence on the precise form 
of the halo profile is never very large though.  It is mostly the 
propagation parameters that determine the shape of the spectrum.

To conclude, the differences between our models and the KT propagation
model are mainly a matter of different parameters.

In Fig.~\ref{fig:epcfb} we show some other typical spectra (examples
1--3 in Table~\ref{tab:examples}). In (a) we show a spectrum when
gauge bosons don't dominate in the final state.  In this case, we
typically get a big wide bump without any special features.  In (b),
we show a spectrum for a heavy neutralino annihilating into gauge
bosons. As in Fig.~\ref{fig:epcfa}, we see two bumps from the decay of
the $W$s. The right one comes from direct decay to $e^{+}$ and occurs
at approximately $m_{\chi}/2$\@. The left one comes from secondary
decays of $W$s, $W^{+} \to \tau^{+}(\mu^{+}) \to e^{+}$ and positrons
from quark jets, $W^+ \to \cdots \to \pi^{+} \to e^+$.  This bump
occurs at approximately $m_{\chi}/60$\@. Exactly where the bumps occur
depend on the propagation model used.
In (c), we show a spectrum of a light neutralino where we have
artificially boosted the branching ratio for direct annihilation into
$e^{+}e^{-}$ to 1\%.  Typically this branching ratio is below
$10^{-7}$, but if we e.g.\ have a large mass splitting of the
selectrons, i.e.\ $m_{\tilde{e}_{1}} \gtrsim m_{\chi}$ and
$m_{\tilde{e}_{2}} \gg m_{\chi}$ we can find mostly gaugino-like, but
also some mixed models with a branching ratio $B_{e^+ e^-} \gtrsim
0.01$\@.  In this case, the line from monochromatic positrons might be
visible and not too widened by propagation.  We find that it is only
when the branching ratio into $e^{+}e^{-}$ directly is reasonably high
($\gtrsim 0.001$) that the line is visible.  We remark that if
selectrons and smuons mix, the selectron degeneracy must be less than
a few percent to avoid constraints from $\mu\rightarrow e\gamma$
\cite{ellisnano,jkg}, though our model is allowed if the mixing
between selectrons and smuons is negligible.

\begin{figure*}[t]
\centerline{\epsfig{file=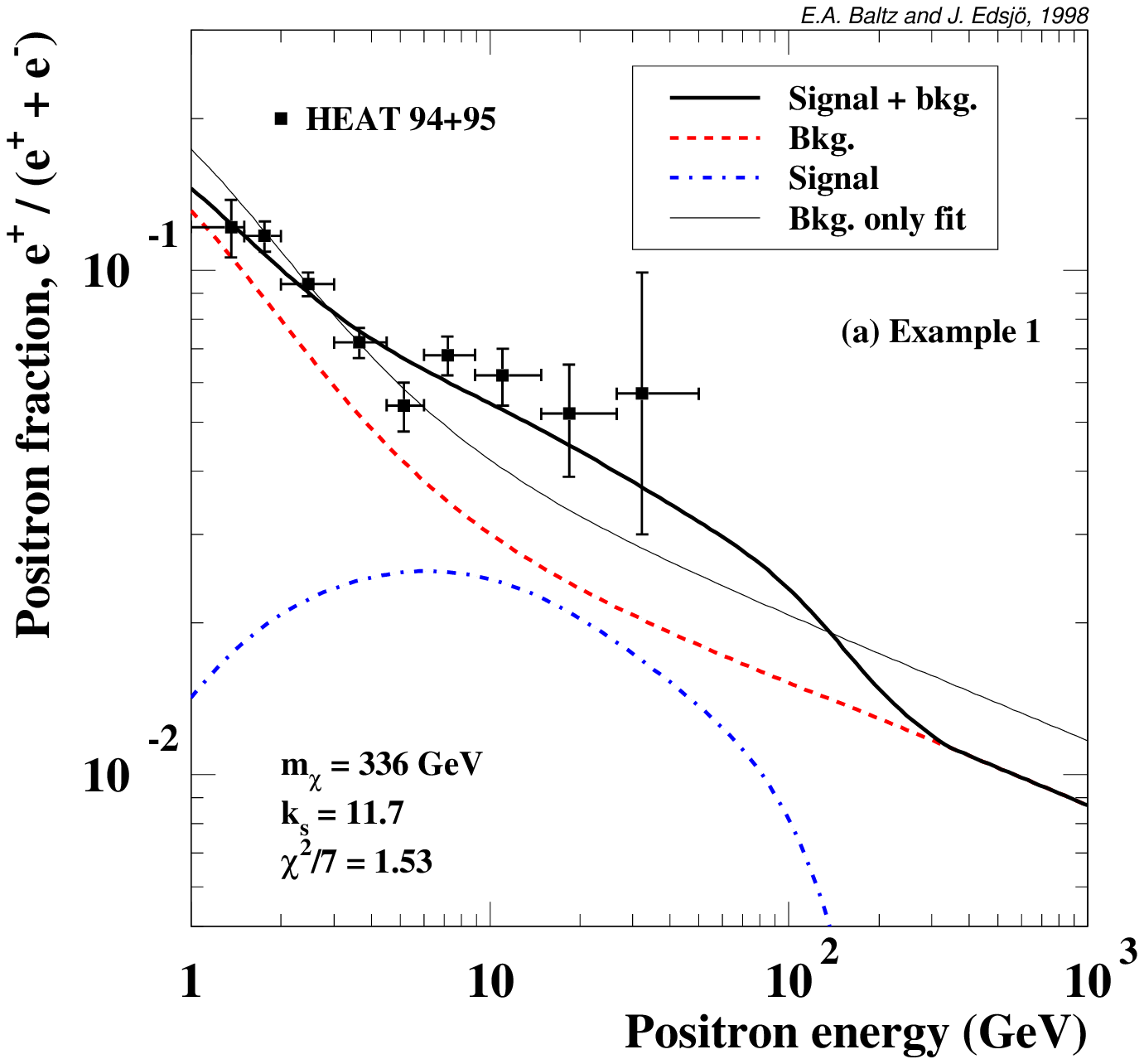,width=0.49\textwidth}
\epsfig{file=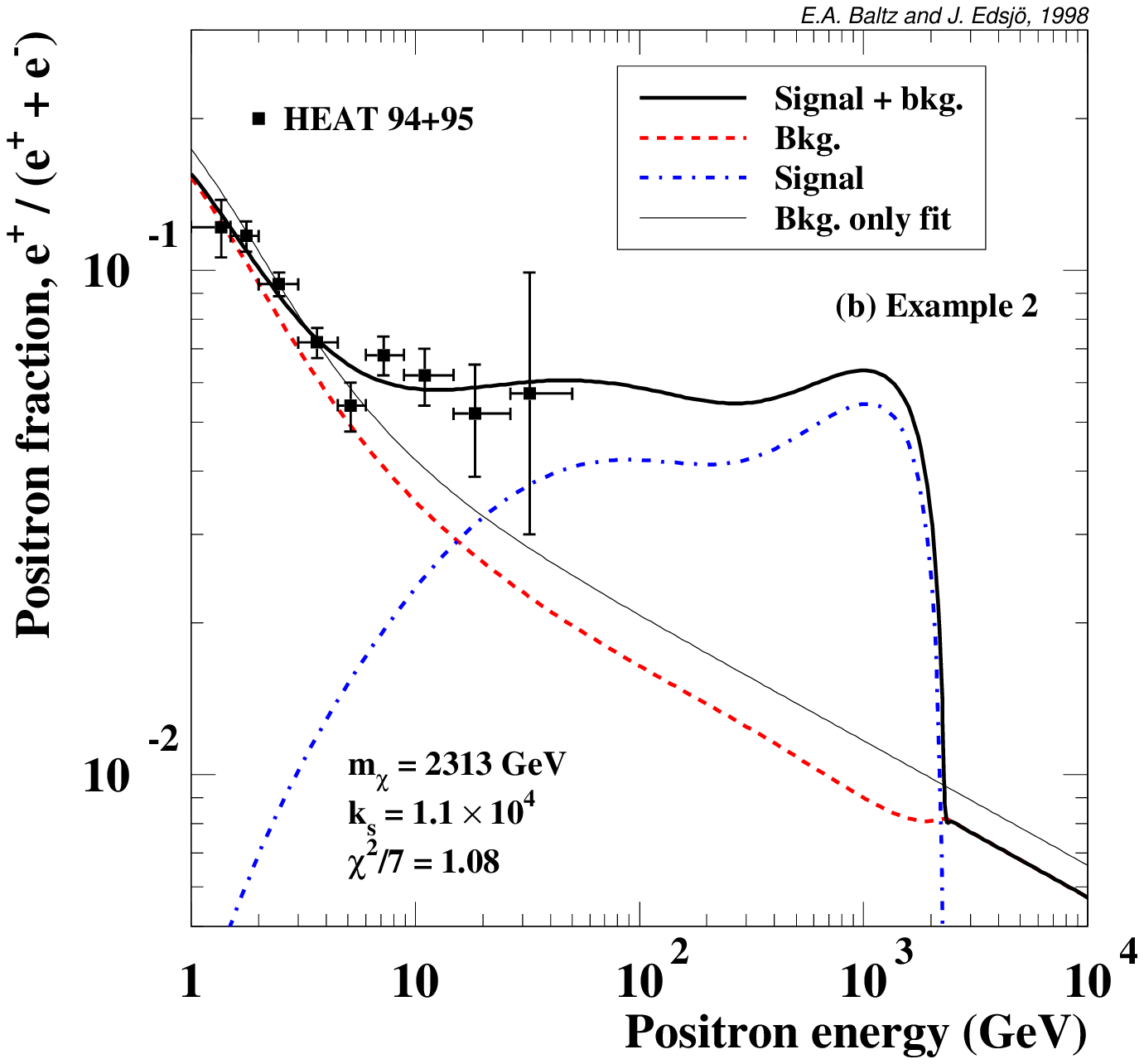,width=0.49\textwidth}}
\centerline{\epsfig{file=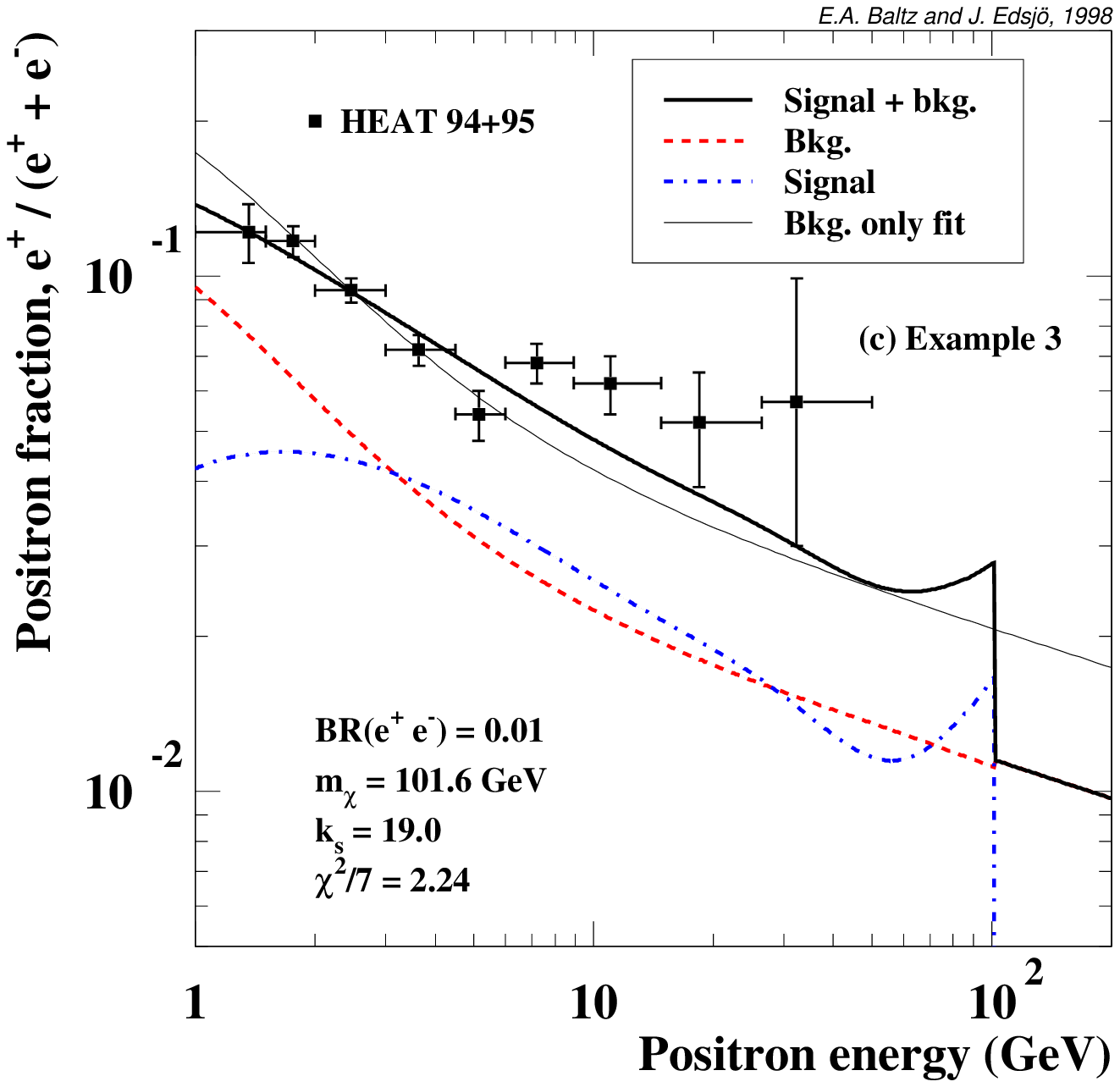,width=0.49\textwidth}
\epsfig{file=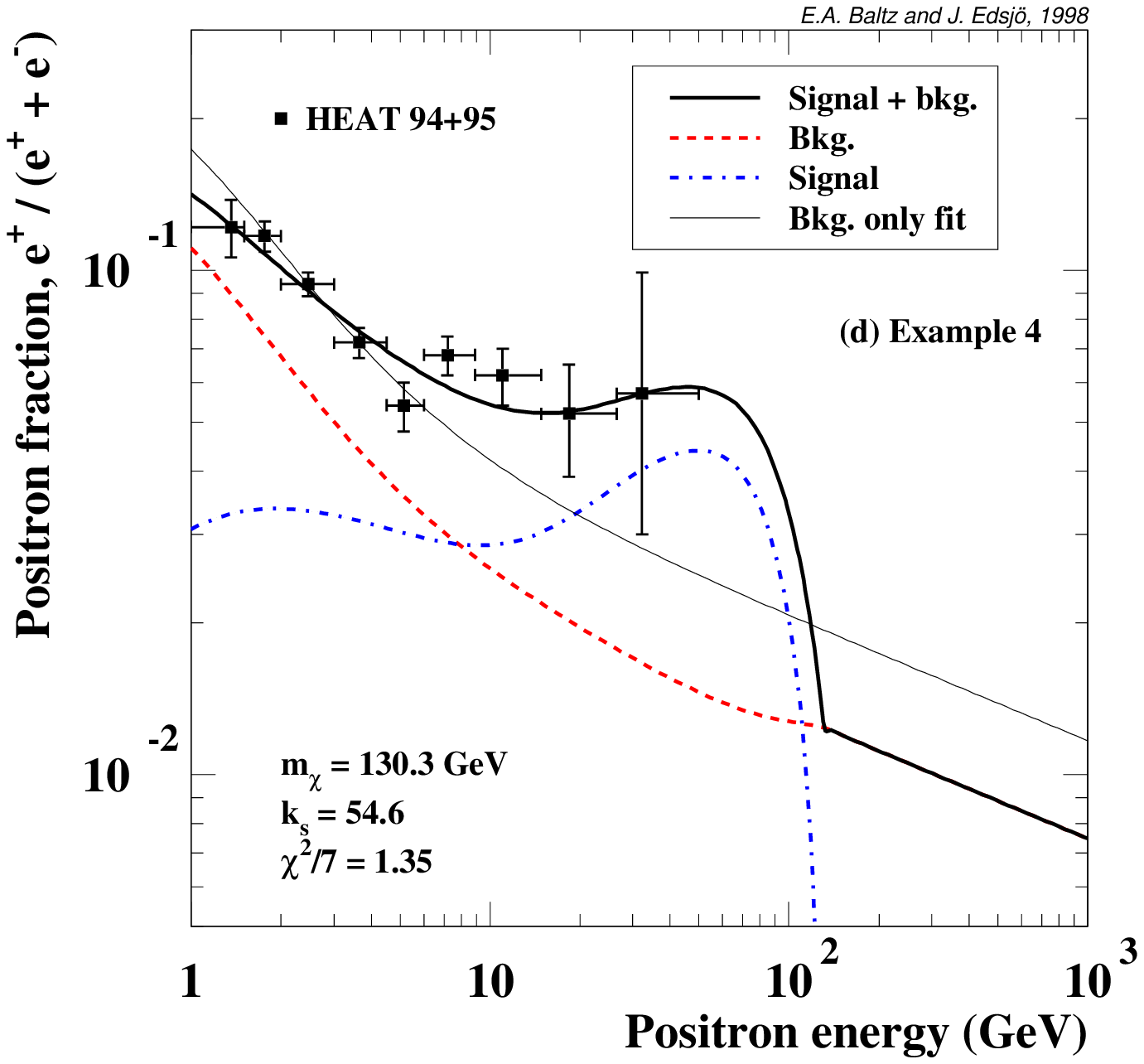,width=0.49\textwidth}} \caption{Example 
of positron fractions versus the positron energy.  We show both the 
signal and background for the best simultaneous fit and also the best 
fit of the background alone.  In (a) we show a typical spectrum for 
medium-heavy neutralinos, in (b) we show a spectrum for a heavy 
neutralino when gauge bosons dominate the annihilation final states, 
in (c) a spectrum when the branching ratio for annihilation into 
$e^{+}e^{-}$ has been increased to 0.01 and in (d) a spectrum from a 
medium-heavy neutralino when $W$s dominate.}
\label{fig:epex}
\end{figure*}

\subsection{Positron fractions}

Instead of comparing with the absolute fluxes, we will now compare
with the measurement of the positron fraction by the {\sc Heat}
experiment \cite{heatfrac}. As we saw in Fig.~\ref{fig:epbkg}, the
standard background predictions fail to reproduce the observations for
all energies. Note, however, that the fit can be made better by
adopting either a harder electron primary spectrum or a harder
interstellar nucelon spectrum \cite{MoskStrong98,MSRdiffgamma}. A
harder electron spectrum seems to be in disagreement with high-energy
electron measurements \cite{MoskStrong98} and a harder nucleon
spectrum might be in contradiction with antiproton measurements
above 3 GeV \cite{MSRdiffgamma}. In light of this, it is therefore
interesting to investigate if the possible excess at 6--50 GeV
in the {\sc Heat} data can be explained by neutralino annihilations in
the halo. Note, however, the the uncertainties in the solar modulation
(even though the charge-dependent solar modulation seems to worsen the
fit) are large and not well understood.

We will in the following entertain the possibility that the standard
prediction of the background is correct (to within a normalization)
and make a simultaneous fit of the normalization of the background and
the signal.  We will not include the charge-dependent solar
modulation.  A study of the {\sc Heat} data in terms of primary
sources from WIMP annihilations has previously been performed in
Ref.~\cite{heatprimary} where the KT results (both with and without
energy-dependent escape time) were used.

We will compare with the most recent {\sc Heat} measurement of the
positron fraction (1994 and 1995 combined data) \cite{heatfrac}.  The
error bars are smaller and the data cleaner for the positron fraction
measurements.  The {\sc Heat} Collaboration gives their results in 9
bins from 1.0 to 50.0 GeV\@.  

As we saw in Fig.~\ref{fig:phiepvsmx} (a), the positron fluxes are
typically an order of magniture or more smaller than the {\sc Heat}
measurements, and we find that we need to boost the signal by a boost
factor, $k_{s}$, between 6 and $10^{10}$ to obtain a good fit.
$k_{s}=10^{10}$ is hardly realistic, but $k_{s}$ up to 100--1000 might
be acceptable given the uncertainties in both propagation and halo
structure (the halo could e.g.\ be clumpy \cite{silkstebbins,clumpy}).

In Fig.~\ref{fig:epex} we show some examples of our positron fractions
compared with the {\sc Heat} measurements.  The examples are the same
as those shown earlier for the absolute fluxes, i.e. examples 1--4 in
Table~\ref{tab:examples}. The signals have been boosted by the
boost factors given by the best fit.  Almost all of our models have
the general feature of increasing the flux at intermediate energies as
shown in (a).  For some models, where the annihilation predominantly
occurs to gauge bosons, we also expect a peak at roughly $m_{\chi}/2$
as shown in (b).  The most pronounced feature, however, is the
positron line.  Even though smeared by propagation and energy loss, we
still expect a sharp peak at $m_{\chi}$ if the branching ratio for
annihilating into monochromatic positrons is high enough.  However,
the branching ratio is typically less than $10^{-7}$ and the feature
is buried in the background.  If the branching ratio is higher, the
feature would be very clear as seen in (c).  This might happen if
there is a large mass splitting of the selectrons as discussed
earlier. In (d) we show a light model annihilating mainly into gauge
bosons and as seen, the bump from primary decay $W^+ \to e^+$ can be
made to fit the excess quite nicely.

In Ref.\ \cite{heatprimary} models were found where the low-energy
bump from gauge bosons can fit the excess at 6--50 GeV (as in our
Fig.~\ref{fig:epex} (c)) without having to go to $\gtrsim$ 1 TeV
neutralinos.  They get better fits for lower masses than we do
because they use the KT propagation model \cite{kamturner} which doesn't
shift features down in energy as much as our propagation does.

\begin{figure*}[t]
\centerline{\epsfig{file=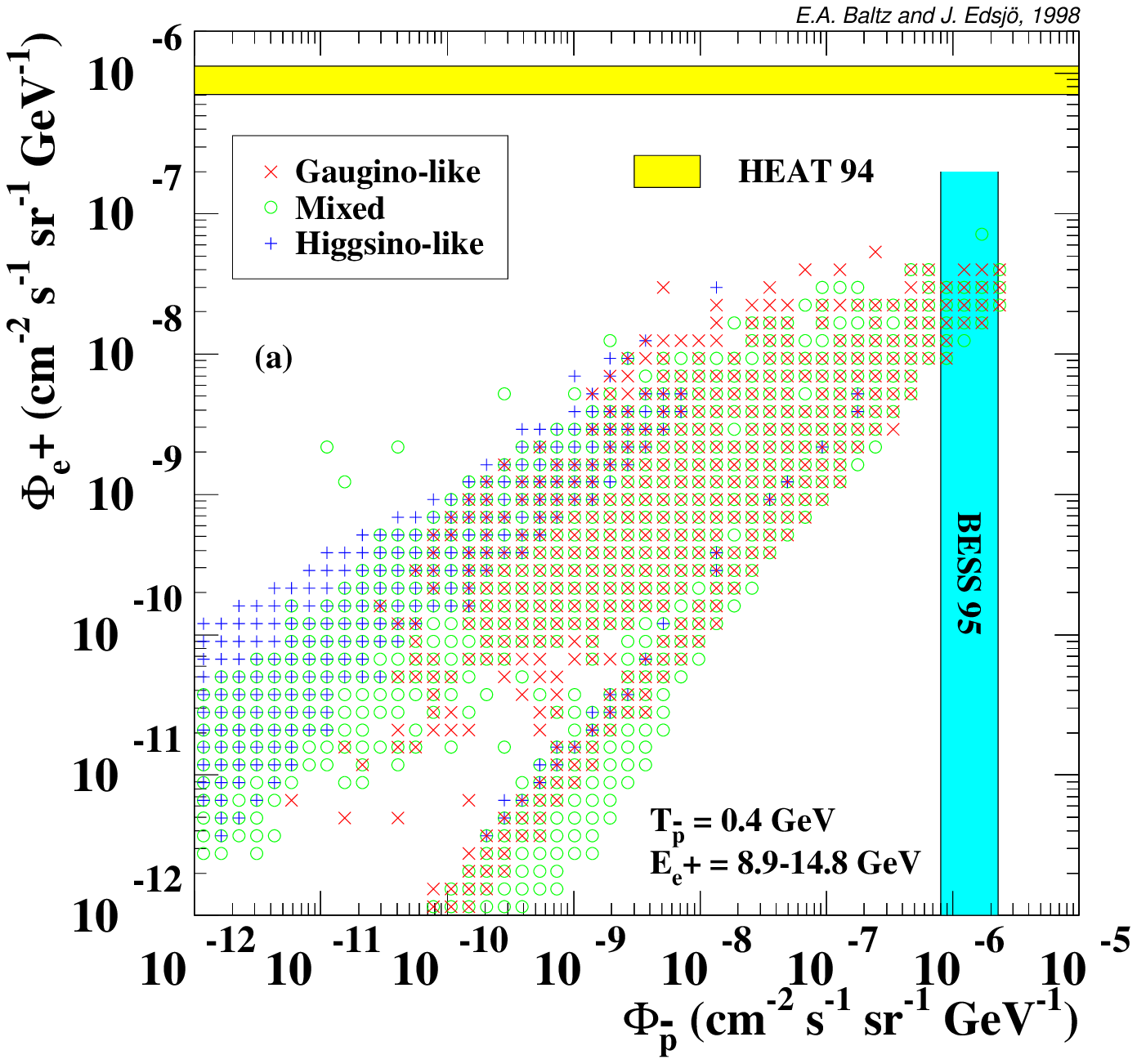,width=0.49\textwidth}
\epsfig{file=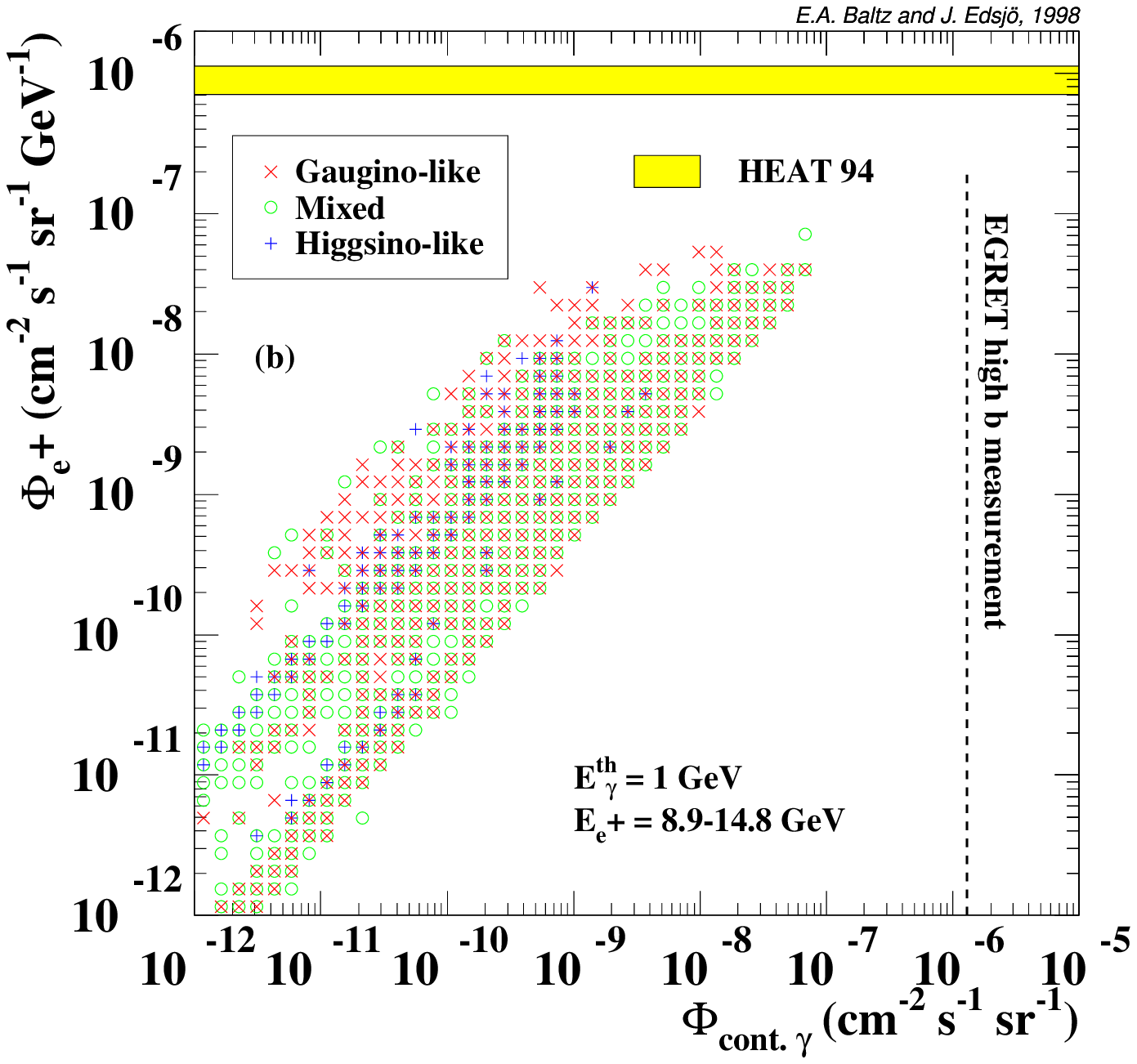,width=0.49\textwidth}} 
\caption{The flux of positrons versus that of (a) antiprotons and (b) 
continuum gammas.  The experimental limits from {\sc Bess} 95 
\protect\cite{bess} and high-altitude {\sc Egret} 
\protect\cite{sreekumar} measurements are shown.}
\label{fig:pbarpgac}
\end{figure*}

\subsection{Comparison with other signals}

We have seen that the positron fluxes are typically at least factors of a few
too low to be observable. However, with only a small amount of boosting of the
signal (coming either from different propagation models and/or clumping of the
dark matter) we obtain positron fluxes that can either roughly fit the
indication of an excess seen in the {\sc Heat} data or produce features
possible to observe with future experiments. We must be concerned with other
signals that might be boosted at a similar level and that might be in conflict
with current observations.

In Fig.~\ref{fig:pbarpgac} we show the absolute flux of positrons
versus the flux of (a) antiprotons \cite{pbarwork,clumpy} and (b)
continuum gammas \cite{clumpy}.  The antiproton measurement of the
{\sc Bess} collaboration \cite{bess} is shown as well as the limit of
the high galactic altitude diffuse gamma emission as measured by {\sc
Egret} \cite{sreekumar}. We see that especially the antiprotons are
expected to give about the same or better constraints on neutralino
dark matter than the positrons. For the models with reasonable boost
factors ($\lesssim$100--1000) the antiproton flux is usually about a
factor of 1--10 closer to the experimental bound than the
positrons. We might therefore worry about getting antiproton fluxes
that are too large when boosting the positron fluxes. However, even
though a factor of 10 might seem to be high, we have large
uncertainties involved in the propagation and solar modulation. For
the positrons we also have large uncertainties from the energy loss
that don't enter in the antiproton propagation. Changes in the
diffusion constant also affect the positron and antiproton fluxes
differently. Increasing $K_0$, the positron fluxes at higher energies
are mainly unchanged (see Fig.~\ref{fig:epcfa} (b)), whereas the
antiproton fluxes decrease.  We also tend to sample a larger volume
for antiprotons than we do for positrons and different halo profiles
and clumpiniess hence enter differently for antiprotons and
positrons. A factor of 10 difference is thus probably within the
uncertainties.

For our examples and boost factors in Fig.~\ref{fig:epex}, the
corresponding maximal boost factors that would not violate the {\sc
Bess} upper limit \cite{bess} of $2.3\times 10^{-6}$ GeV$^{-1}$
cm$^{-2}$ s$^{-1}$ sr$^{-1}$ on the antiproton flux would be $k_{\bar
p}$ = 9.8, $3.5 \times 10^5$, 3.2 and 10.7.  Compare this with the
boost factors from our best fits to the {\sc Heat} data, $k_s$ = 11.7,
$1.1 \times 10^4$, 19.0 and 54.6.  The last two examples,
Fig.~\ref{fig:epex} (c) and (d) have boost factors which contradicts
the antiproton measurements by a factor of 5--6 but as explained
above, the uncertainties are large and we prefer to keep these kind of
models anyway.

The continuum gammas are very well correlated with the positron fluxes 
and of the same order of magnitude below present limits.  The 
continuum gamma fluxes don't suffer from large propagation 
uncertainties, but do depend more on the actual halo profile.

To conclude, it is intriguing that both the positron, antiproton and 
continuum gamma fluxes for our canonical halo and propagation models 
end up reasonably close to the observed fluxes.  The correlation is 
also quite good, even though different uncertainties enter in all 
three different predictions.

Other signals, like direct detection of neutralinos and indirect detection 
from high-energy neutrinos from annihilation in the Sun/Earth 
are not affected by clumpiness to the same extent as the ones from 
halo annihilation.  Both direct searches and neutrino telescopes have 
started cutting into the MSSM parameter space considered here.  
However, the limits are not completely watertight and we choose to keep 
all models even though some of our models might be excluded with 
typical assumptions about the halo density and velocity dispersion.

\section{Conclusions}

We conclude that for our canonical halo and propagation model, the
positron fluxes from neutralino annihilation in the halo are lower
than those experimentally measured. However, the predictions can
easily be orders of magnitude too low due to our lack of knowledge of
the structure of the halo (the halo could e.g.\ be clumpy).  If we
allow for this uncertainty we can easily obtain fluxes of the same
order of magnitude as those measured. The shape of the signal spectrum
is also different than the background and in most cases allows for a
better fit of the excess at 6--50 GeV indicated by the {\sc Heat}
measurements than the background alone does. 

If positrons from neutralinos make up a substantial fraction of the
measured positron fluxes below 50 GeV, we usually have some features
to search for at higher energies, especially if the annihilation
occurs mainly to gauge bosons where we expect a clear bump at
$m_\chi/2$\@.  When the annihilation doesn't go to gauge bosons, we
only excpect a slight break in the spectrum at approximately
$m_\chi/2$\@. This can be hard to find, unless we have detailed
measurements around the break. If the branching ratio for annihilation
into monochromatic positrons is higher than $\gtrsim 0.001$, as can
happen if there is a pronounced mass splitting between selectrons,
there is a very pronounced feature to search for at a positron energy
of $m_\chi$\@. 

We also note that our propagation models don't preserve features as
well as that of KT \cite{kamturner}.  This is primarily due to the
fact that KT uses an escape time that implies a larger diffusion
constant.  Features are also shifted to lower energies with our
propagation models.

To conclude, there is a possibility to obtain measureable fluxes of
positrons from neutralino annihilation in the Milky Way halo and they
can easily be made to fit the excess indicated by {\sc Heat} data.
Future experiments will determine if there are features in the
positron spectrum from neutralino annihilation.

\acknowledgments

We wish to thank G.~Tarl{\'e} for useful discussions on the {\sc Heat}
data.  We also thank P.~Gondolo for discussions at an early stage of
this work. This work was supported with computing resources by the
Swedish Council for High Performance Computing (HPDR) and
Parallelldatorcentrum (PDC), Royal Institute of Technology.  E.~B.\
was supported by grants from NASA and DOE. J.~E.\ was supported by an
Uppsala-Berkeley exchange program from Uppsala University.


\end{document}